\pdfoutput=1

\documentclass[letterpaper,twocolumn,10pt]{article}

\usepackage{usenix2019_v3} %,epsfig}
\usepackage{mdframed}

\usepackage{enumitem}
\usepackage[compact,small]{titlesec}
\usepackage{hhline}
\usepackage{framed}
\usepackage{multirow}
\usepackage{color, colortbl}
\definecolor{LightCyan}{rgb}{0.85,0.85,0.85}
\definecolor{DarkCyan}{rgb}{0.9,0.9,0.9}
\definecolor{DarkGreen}{rgb}{0,0.5,0}
\definecolor{LightGreen}{rgb}{0.8,0.95,0.8}
\definecolor{LightRed}{rgb}{0.95,0.8,0.8}
\usepackage{booktabs}

\usepackage{amssymb}
\usepackage{pifont}
\usepackage{amsmath}

\usepackage{subfigure}
\usepackage{algorithm}
\usepackage{setspace}
\usepackage[linesnumbered,algo2e,ruled]{algorithm2e}

\makeatletter
\def\mathcolor#1#{\@mathcolor{#1}}
\def\@mathcolor#1#2#3{%
  \protect\leavevmode
  \begingroup
    \color#1{#2}#3%
  \endgroup
}
\makeatother

\usepackage{xfrac}
\usepackage{footnote}

\makesavenoteenv{tabular}

\usepackage{xargs}
\usepackage{url}

\usepackage{graphicx}
\usepackage{listings}
\newcommand{\pnt}{\emph{pnt}}
\newcommand{\cum}{\emph{cml}}

\newcommand{\dataset}{$D$}
\newcommand{\labelingcost}{$L$}
\newcommand{\quarantinecost}{$Q$}
\newcommand{\performance}{$P$}
\newcommand{\performancestar}{$P^*$}

\newcommand{\errorrate}{$E$}
\newcommand{\errorratemax}{$E_{max}$}

\newcommand{\drebin}{\textsc{Alg1}}
\newcommand{\mmd}{\textsc{Alg2}}
\newcommand{\dl}{\textsc{DL}}

\newcommand{\sklearn}{\textsf{\textsc{Scikit-learn}}}
\newcommand{\keras}{\textsf{\textsc{Keras}}}

\newcommand{\tool}{\textsc{Tesseract}}

\usepackage{hyperref} % not needed; should go before cleveref if loaded
\hypersetup{colorlinks,%
filecolor=black,%
linkcolor=black,%
urlcolor=black
}

\usepackage{cleveref}

\def\Snospace~{\S{}}

\begin{document}

%don't want date printed
\date{}

\title{TESSERACT: Eliminating Experimental Bias in Malware Classification\\across Space and Time}

\author{
{\rm Feargus~Pendlebury\footnotemark[1]~\footnotemark[2]~\footnotemark[3]~,  Fabio~Pierazzi\footnotemark[1]~\footnotemark[2]~\footnotemark[3]~, Roberto~Jordaney\footnotemark[2]~\footnotemark[3]~, Johannes~Kinder\footnotemark[4]~, Lorenzo~Cavallaro\footnotemark[2]}\\
\footnotemark[2]~~King's College London\\
\footnotemark[3]~~Royal Holloway, University of London\\
\footnotemark[4]~~Bundeswehr University Munich
} % end author

\maketitle

% Forcing just the first symbol as footnote
\renewcommand*{\thefootnote}{\fnsymbol{footnote}}
\footnotetext[1]{Equal contribution.}
\setcounter{footnote}{0}
\renewcommand*{\thefootnote}{\arabic{footnote}}

\begin{abstract}
Is Android malware classification a solved problem? Published $F_1$
scores of up to 0.99 appear to leave very little room for improvement.
In this paper, we argue that results are commonly inflated due to two
pervasive sources of experimental bias: \textit{spatial bias} caused
by distributions of training and testing data that are not
representative of a real-world deployment; and \textit{temporal bias}
caused by incorrect time splits of training and testing
sets, leading to impossible configurations.  We propose a set of space
and time constraints for experiment design that eliminates both sources
of bias. We introduce a new metric that summarizes the expected
robustness of a classifier in a real-world setting, and we present an
algorithm to tune its performance. Finally, we demonstrate how this
allows us to evaluate mitigation strategies for time decay such as
active learning.  We have implemented our solutions in \tool{}, an
open source evaluation framework for comparing
malware classifiers in a realistic setting. We used \tool{} to
evaluate three Android malware classifiers from the literature on a
dataset of 129K applications spanning over three years. Our evaluation
confirms that earlier published results are biased, while also revealing
counter-intuitive performance and showing that appropriate tuning can
lead to significant improvements.
\end{abstract}

\section{Introduction}

Machine learning has become a standard tool for malware research in
the academic security community: it has been used in a wide range of
domains including
Windows malware~\cite{Dahl:Large,Tahan:Mal,Markel:Building},
PDF malware~\cite{Laskov:PDF,Maiorca:PDF},
malicious URLs~\cite{Stringhini:Shady,Lee:Warningbird},
malicious JavaScript~\cite{Rieck:Cujo,Curtsinger:Zozzle},
and Android malware~\cite{Arp:Drebin,Mariconti:MaMaDroid,Papernot:ESORICS}.
With tantalizingly high performance figures, it seems malware should be a problem of the past.

Malware classifiers operate in dynamic contexts. As malware evolves and new variants and families appear over time, prediction quality decays~\cite{Jordaney:Transcend}. Therefore, temporal consistency matters for evaluating the effectiveness of a classifier. When the experimental setup allows a classifier to train on what is effectively future knowledge, the reported results become biased~\cite{Allix:Timeline, Miller:Reviewer}.

This issue is widespread in the security community and affects multiple security domains. In this paper, we focus on Android malware and claim that there is an endemic issue in that Android malware classifiers~\cite{Arp:Drebin,Suarez:DroidSieve,Mariconti:MaMaDroid,Gascon:Structural,Zhang:Semantics,Dash:Droidscribe,Yuan:DroidSec,Papernot:ESORICS} (including our own work) are not evaluated in settings representative of real-world deployments.
We choose Android because of the availability of (a) a public, large-scale, and timestamped dataset
(AndroZoo~\cite{Allix:AndroZoo}) and (b) algorithms that are feasible to
reproduce (where all~\cite{Mariconti:MaMaDroid} or
part~\cite{Arp:Drebin} of the code has been released).

We identify experimental bias in two dimensions, \emph{space} and \emph{time}.
\textit{Spatial bias} refers to unrealistic assumptions about the ratio of goodware to malware in the data. The ratio of goodware to malware is domain-specific, but it must be enforced consistently during the testing phase to mimic a realistic scenario. For example, measurement studies on Android suggest that most apps in the wild are goodware~\cite{lindorfer2014andradar,Google:Report}, whereas for (desktop) software download events most URLs are malicious~\cite{Maggi:URLs, Perdisci:URLs}.
\textit{Temporal bias} refers to temporally inconsistent evaluations which integrate future knowledge about the testing objects into the training phase~\cite{Allix:Timeline,Miller:Reviewer} or create unrealistic settings. This problem is exacerbated by families of closely related malware, where including even one variant in the training set may allow the algorithm to identify many variants in the testing.

We believe that the pervasiveness of these issues is due to two main
reasons: first, possible sources of evaluation bias are not common
knowledge; second, accounting for time complicates the evaluation and
does not allow a comparison to other approaches using headline
evaluation metrics such as the $F_1$-Score or AUROC. We address these
issues in this paper by systematizing evaluation bias for Android
malware classification and providing new constraints for sound
experiment design along with new metrics and tool support.

Prior work has investigated challenges and experimental bias in security evaluations~\cite{rossow2012prudent,Sommer:Outside,Van:BenchmarkingCrimes, Allix:Timeline, Miller:Reviewer, Axelsson:BaseRate}. The \emph{base-rate fallacy}~\cite{Axelsson:BaseRate} describes how evaluation metrics such as $\mathit{TPR}$ and $\mathit{FPR}$ are misleading in intrusion detection, due to significant class imbalance (most traffic is benign); in contrast, we identify and address experimental settings that give misleading results \emph{regardless} of the adopted metrics---even when correct metrics are reported.
Sommer and Paxson~\cite{Sommer:Outside}, Rossow et
al.~\cite{rossow2012prudent}, and van der Kouwe et
al.~\cite{Van:BenchmarkingCrimes} discuss possible guidelines for
sound security evaluations; but none of these works identify temporal
and spatial bias, nor do they \emph{quantify} the impact of errors on
classifier performance. Allix et al.~\cite{Allix:Timeline} and
Miller et al.~\cite{Miller:Reviewer} identify an initial temporal
constraint in Android malware classification, but we show that even
results of recent work following their guidelines
(e.g.,~\cite{Mariconti:MaMaDroid}) suffer from other temporal and
spatial bias (\autoref{sec:theexample}). To the best of our knowledge, we are the first to identify and address these sources of bias with novel, actionable constraints, metrics, and tool support~(\autoref{sec:rules}).

This paper makes the following contributions:

\begin{itemize}  [noitemsep, nolistsep]

\item We identify \emph{temporal} bias associated with incorrect train-test   splits~(\autoref{sec:bias-time}) and \emph{spatial} bias related to unrealistic assumptions in dataset distribution~(\autoref{sec:bias-space}). We experimentally verify on a dataset of 129K apps (with 10\% malware) that, due to bias, performance can decrease up to $50\%$ in practice~(\autoref{sec:problem}) in two well-known Android malware classifiers, \textsc{Drebin}~\cite{Arp:Drebin} and \textsc{MaMaDroid}~\cite{Mariconti:MaMaDroid}, which we refer to as \drebin{} and \mmd{}, respectively.

\item We propose novel building blocks for more robust evaluations of
malware  classifiers: a set of spatio-temporal constraints to be enforced in experimental  settings (\autoref{sec:constraints}); a new metric, AUT, that captures a  classifier's robustness to time decay in a single number and allows for the  fair comparison of different algorithms (\autoref{sec:aut}); and a novel tuning algorithm that empirically optimizes the classification performance, when malware represents the minority class (\autoref{sec:tuning}). We compare the performance of \drebin{}~\cite{Arp:Drebin}, \mmd{}~\cite{Mariconti:MaMaDroid} and \dl{}~\cite{Papernot:ESORICS} (a deep learning-based approach), and show how removing bias can provide counter-intuitive results on real performance~(\autoref{sec:theexample}).

\item We implement and publicly release the code of our
methodology (\autoref{sec:rules}), \tool{}, and we further demonstrate
how our findings can be used to evaluate performance-cost trade-offs
of solutions to mitigate time decay such as active
learning~(\autoref{sec:delay}).

\end{itemize}

\tool{} can assist the research community in producing comparable
results, revealing counter-intuitive performance, and assessing a classifier's
prediction qualities in an industrial deployment (\autoref{sec:discussion}).

We believe that our methodology also creates an opportunity to evaluate the
extent to which spatio-temporal experimental bias affects
security domains other than Android malware, and we encourage the security community to embrace
its underpinning philosophy.

\textbf{Use of the term ``bias'':} We use \emph{(experimental) bias} to refer
  to the details of an experimental setting that depart from the conditions in a
  real-world deployment and can have a misleading impact (\emph{bias}) on
  evaluations. We do not intend it to relate to the classifier bias/variance
  trade-off~\cite{Bishop:ML} from traditional machine learning terminology.

\section{Android Malware Classification}
\label{sec:casestudy}

We focus on Android malware classification. In~\autoref{sec:algo} we introduce the reference approaches evaluated, in~\autoref{sec:tsdist} we discuss the domain-specific prevalence of malware, and in~\autoref{sec:dataset} we introduce the dataset used throughout the paper.

\subsection{Reference Algorithms}
\label{sec:algo}

To assess experimental bias (\autoref{sec:bias}), we consider two high-profile machine learning-driven techniques for Android malware classification, both published recently in top-tier security conferences.
The first approach is {\bf\drebin{}}~\cite{Arp:Drebin}, a linear support vector machine (SVM) on high-dimensional binary feature vectors engineered with a lightweight static analysis.
The second approach is {\bf\mmd{}}~\cite{Mariconti:MaMaDroid}, a Random Forest (RF) applied to features engineered by modeling caller-callee relationships over Android API methods as Markov chains.
We choose \drebin{} and \mmd{} as they build on different types of
static analysis to generate feature spaces capturing Android
application characteristics at different levels of abstraction;
furthermore, they use different machine learning algorithms to learn
decision boundaries between benign and malicious Android apps in the
given feature space. Thus, they represent a broad design space and
support the generality of our methodology for characterizing
experimental bias.  For a sound experimental baseline, we
reimplemented \drebin{} following the detailed description in the
paper; for \mmd{}, we relied on the implementation provided by its
authors. We replicated the baseline results for both
approaches.
After identifying and quantifying the impact of experimental bias (\autoref{sec:bias}), we propose specific constraints and metrics to allow fair and unbiased comparisons (\autoref{sec:rules}). Since \drebin{} and
\mmd{} adopt traditional ML algorithms, in~\autoref{sec:rules} we also consider {\bf
\dl{}}~\cite{Papernot:ESORICS}, a deep learning-based approach that
takes as input the same features as \drebin{}~\cite{Arp:Drebin}. We
include \dl{} because the latent feature space of deep
learning approaches can capture different representations of the input
data~\cite{Goodfellow:DL}, which may affect their robustness to time decay. We replicate the baseline results for \dl{} reported in~\cite{Papernot:ESORICS} by re-implementing its neural network architecture and by using the same input features as for \drebin{}.

It speaks to the scientific standards of these papers that we were
able to replicate the experiments; indeed, we would like to
emphasize that we do not criticize them specifically. We use these
approaches for our evaluation because they are available and offer stable
baselines.

We report details on the hyperparameters of the reimplemented algorithms in Appendix~\ref{sec:hyperparam}.

\subsection{Estimating in-the-wild Malware Ratio}
\label{sec:tsdist}

The proportion of malware in the dataset can greatly affect the
performance of the classifier (\autoref{sec:bias}).  Hence, unbiased
experiments require a dataset with a realistic percentage of malware
over goodware; on an already existing dataset, one may enforce such a
ratio by, for instance, downsampling the majority class (\autoref{sec:bias-space}).
Each malware domain has its own, often unique, ratio of malware to goodware typically
encountered in the wild. First, it is important to know if malware is
a minority, majority, or an equal-size class as goodware. For example,
malware is the minority class in network
traffic~\cite{Axelsson:BaseRate} and
Android~\cite{lindorfer2014andradar}, but it is the majority class in
binary download events~\cite{Perdisci:URLs}.
On the one hand, the estimation of the percentage of malware
in the wild for a given domain is a non-trivial task. On the
other hand, measurement papers, industry telemetry, and
publicly-available reports may all be leveraged to obtain realistic
estimates.

In the Android landscape, malware represents 6\%--18.8\% of all the
apps, according to different sources: a key industrial
player{}\footnote{Information obtained through confidential emails with the authors.}
reported the ratio as approximately 6\%, whereas the AndRadar measurement
study~\cite{lindorfer2014andradar} reports around 8\% of Android
malware in the wild. The 2017 Google's Android security
report~\cite{Google:Report} suggests 6--10\% malware, whereas an
analysis of the metadata of the AndroZoo dataset~\cite{Allix:AndroZoo}
totaling almost 8M Android apps updated regularly, reveals an
incidence of 18.8\% of malicious apps.
The data suggests that, in the Android domain, malware is the
minority class. In this work, we decide to stabilize its percentage
to 10\% (a de-facto average across the various
estimates), with per-month values between 8\% and 12\%. Settling on an
average overall ratio of 10\% Android malware also allows us to
collect a dataset with a statistically sound number of per-month
malware. An aggressive undersampling would have
decreased the statistical significance of the dataset, whereas
oversampling goodware would have been too resource intensive
(\autoref{sec:dataset}).

\subsection{Dataset}
\label{sec:dataset}

We consider samples from the public AndroZoo~\cite{Allix:AndroZoo} dataset, consisting of  more than 8.5 million Android apps between 2010 and early 2019: each app is associated with a timestamp, and most apps include VirusTotal metadata results.
The dataset is constantly updated by crawling from different markets (e.g., more than 4 million apps from Google Play Store, and the remaining from markets such as Anzhi and AppChina). We choose to refer to this dataset due to its size and timespan, which allow us to perform realistic space- and time-aware experiments.

{\bf Goodware and malware.} AndroZoo's metadata reports the number $p$
of positive anti-virus reports on VirusTotal~\cite{VT:URL} for applications in
the AndroZoo dataset. We chose $p=0$ for goodware and $p \geq 4$ for malware, following Miller et al.'s~\cite{Miller:Reviewer} advice for a reliable ground-truth. About 13\% of AndroZoo apps can be called grayware as they have $0 < p < 4$. We exclude grayware from the sampling as including it as either goodware or malware could disadvantage classifiers whose features were designed with a different labeling threshold.

{\bf Choosing apps.}
The number of objects we consider in our study is affected by the feature
extraction cost, and partly by storage space requirements (as the full AndroZoo
dataset, at the time of writing, is more than 50TB of apps to which one must add
the space required for extracting features).
Extracting features for the whole AndroZoo dataset may
take up to three years on our research infrastructure (three high-spec
Dell PowerEdge R730 nodes, each with 2 x 14 cores in hyperthreading---in total, 168 vCPU threads, 1.2TB of RAM, and a 100TB NAS), thus we decided to extract features from
129K apps (\autoref{sec:tsdist}). We believe this represents a large dataset
with enough statistical significance.
To evaluate time decay, we decide on a granularity of one month, and we
uniformly sample 129K AndroZoo apps in the period from Jan 2014 to Dec 2016, but
also enforce an overall average of $10\%$ malware (see \autoref{sec:tsdist})--with an allowed percentage of malware per month between 8\% and 12\%, to ensure some  variability. Spanning over
three years ensures 1,000+ apps per month (except for the last three
months, where AndroZoo had crawled less applications). We consider apps up to Dec 2016 because the VirusTotal results for 2017 and 2018 apps were mostly unavailable from AndroZoo at the time of writing; moreover, Miller et al.~\cite{Miller:Reviewer}
empirically evaluated that antivirus detections become stable after approximately one
year---choosing Dec 2016 as the finishing time ensures good ground-truth
confidence in objects labeled as malware.

\begin{figure}[t]
 	\includegraphics[width=1\columnwidth]{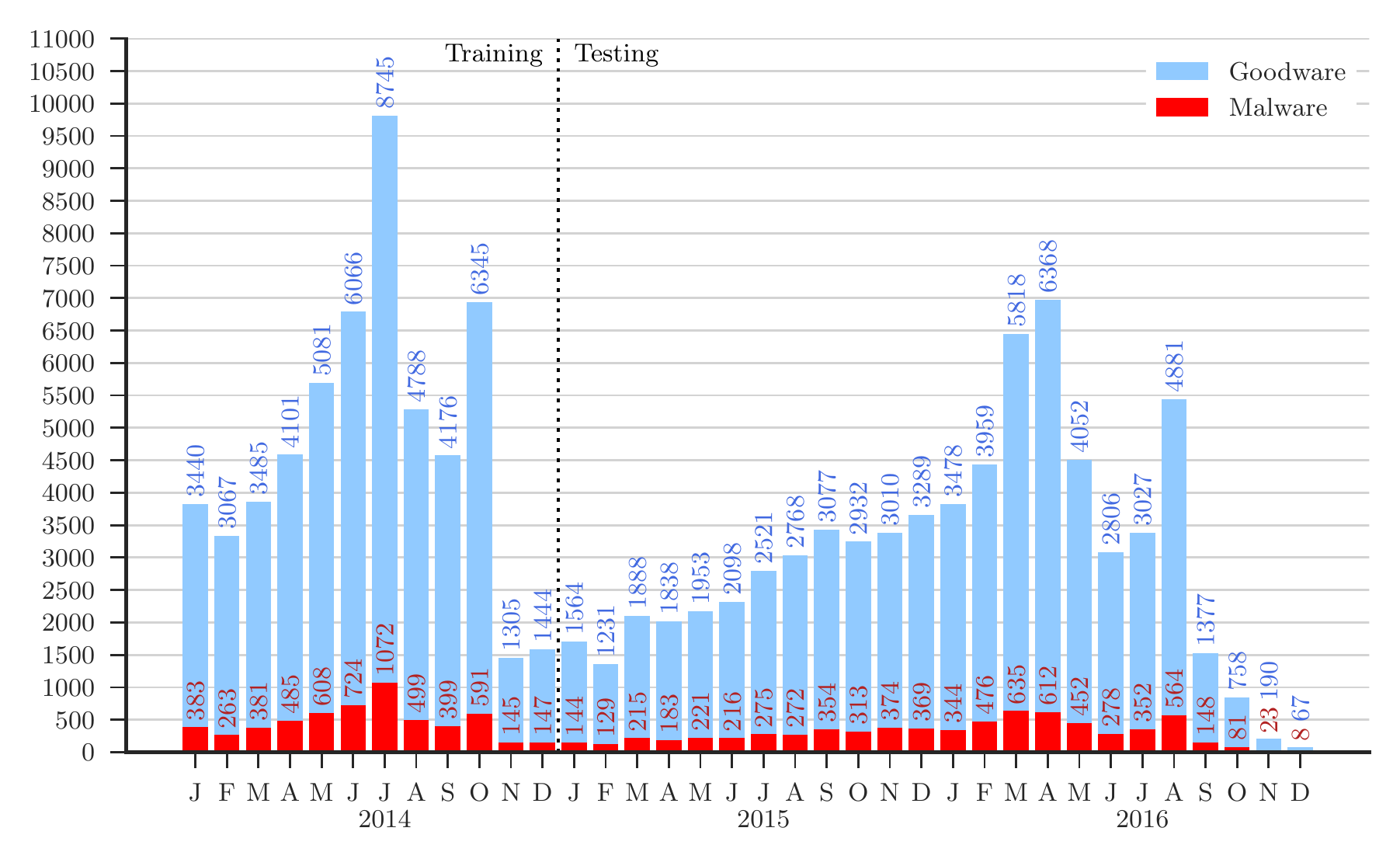}
 	\vspace{-15px}
 	\caption{\emph{Details of the dataset considered throughout this paper}. The figure reports a stack histogram with the monthly distribution of apps we collect from AndroZoo: $129,728$ Android applications (with average 10\% malware), spanning from Jan 2014 to Dec 2016. The vertical dotted line denotes the split we use in all time-aware experiments in this paper (see \autoref{sec:rules} and \autoref{sec:delay}): training on 2014, testing on 2015 and 2016.}
 	\label{fig:dataset}
\end{figure}

{\bf Dataset summary.} The final dataset consists of 129,728 Android applications (116,993 goodware and 12,735 malware). \autoref{fig:dataset} reports a stack histogram showing the per-month distribution of goodware/malware in the dataset. For the sake of clarity, the figure also reports the number of malware and goodware in each bin. The training and testing splits used in \autoref{sec:bias} are reported in \autoref{tab:bias}; all the time-aware experiments in the remainder of this paper are performed by training on 2014 and testing on 2015 and 2016 (see the vertical dotted line in \autoref{fig:dataset}).

    \begin{table*}[t]
        \centering

        \resizebox{0.8\textwidth}{!}{%

        \begin{tabular}{|r||c|c||c|c|c|c||c|c|c|c|}
			\hhline{~~~--------}

             \multicolumn{3}{c||}{} & \multicolumn{8}{c|}{{\bf
\cellcolor{LightCyan} \% mw in testing set Ts}}\\
			\hhline{~~~--------}

             \multicolumn{1}{c}{} & \multicolumn{1}{c}{} & & \multicolumn{4}{c||}{10\% (\emph{realistic})}  & \multicolumn{4}{c|}{90\% (\emph{unrealistic})} \\

			\hhline{~~~--------}
            \multicolumn{1}{c}{} & \multicolumn{1}{c}{} & & \multicolumn{4}{c||}{\bf \cellcolor{LightCyan} \% mw in training set Tr} & \multicolumn{4}{c|}{\cellcolor{LightCyan} \bf \% mw in training set Tr} \\

			\hhline{~----------}
            \multicolumn{1}{c}{} &  \multicolumn{2}{|c||}{\cellcolor{LightCyan}  {\bf Sample dates}}  & 10\% & 90\% & 10\% & 90\% & 10\% & 90\% & 10\% & 90\% \\

			\hline
            \cellcolor{LightCyan} {\bf Experimental setting} &  \multirow{1}{*}{Training} &  \multirow{1}{*}{Testing}  & \multicolumn{2}{c|}{\cellcolor{DarkCyan} \drebin{}~\cite{Arp:Drebin}} & \multicolumn{2}{c||}{\cellcolor{DarkCyan}  \mmd{}~\cite{Mariconti:MaMaDroid}} & \multicolumn{2}{c|}{\cellcolor{DarkCyan}  \drebin{}~\cite{Arp:Drebin}} & \multicolumn{2}{c|}{\cellcolor{DarkCyan} \mmd{}~\cite{Mariconti:MaMaDroid}} \\
		\hline \hline

 \multirow{2}{*}{{\bf 10-fold CV}} & {\tt gw}:  $\blacksquare \blacksquare \blacksquare \blacksquare \blacksquare \blacksquare$  & {\tt gw}:  $\blacksquare \blacksquare \blacksquare \blacksquare \blacksquare \blacksquare $ &  \cellcolor{LightRed}  & \multirow{2}{*}{0.56} & \multirow{2}{*}{0.83}  & \multirow{2}{*}{0.32} & \multirow{2}{*}{0.94} & \multirow{2}{*}{0.98} & \multirow{2}{*}{0.85} & \cellcolor{LightRed}   \\
   & {\tt mw}: $\blacksquare \blacksquare \blacksquare \blacksquare \blacksquare \blacksquare$  & {\tt mw}: $\blacksquare \blacksquare \blacksquare \blacksquare \blacksquare \blacksquare $  &  \cellcolor{LightRed} \multirow{-2}{*}{\bf 0.91} & & & & & & & \cellcolor{LightRed} \multirow{-2}{*}{ \bf 0.97}   \\
            \hline

\multirow{2}{*}{{\bf Temporally inconsistent}} & {\tt gw}: $\mathcolor{lightgray}{\blacksquare}  \mathcolor{lightgray}{\blacksquare}  \mathcolor{lightgray}{\blacksquare} \mathcolor{lightgray}{\blacksquare} \blacksquare  \blacksquare$  & {\tt gw}: $\blacksquare \blacksquare \blacksquare \blacksquare \mathcolor{lightgray}{\blacksquare} \mathcolor{lightgray}{\blacksquare} $  & \multirow{2}{*}{0.76}  & \multirow{2}{*}{0.42} & \multirow{2}{*}{0.49} & \multirow{2}{*}{0.21} & \multirow{2}{*}{0.86} & \multirow{2}{*}{0.93} & \multirow{2}{*}{0.54} & \cellcolor{LightRed}  \\
                     & {\tt mw}: $\mathcolor{lightgray}{\blacksquare}  \mathcolor{lightgray}{\blacksquare}  \mathcolor{lightgray}{\blacksquare} \mathcolor{lightgray}{\blacksquare} \blacksquare  \blacksquare$  & {\tt mw}: $\blacksquare \blacksquare \blacksquare \blacksquare \mathcolor{lightgray}{\blacksquare} \mathcolor{lightgray}{\blacksquare} $  &  & &  & & & & & \cellcolor{LightRed} \multirow{-2}{*}{\bf 0.95} \\
            \hline

 \multirow{1}{*}{{\bf Temporally inconsistent}}& {\tt gw}: $\mathcolor{lightgray}{\blacksquare}  \mathcolor{lightgray}{\blacksquare}  \mathcolor{lightgray}{\blacksquare} \mathcolor{lightgray}{\blacksquare} \blacksquare  \mathcolor{lightgray}{\blacksquare}$  & {\tt gw}: $\mathcolor{lightgray}{\blacksquare} \mathcolor{lightgray}{\blacksquare} \mathcolor{lightgray}{\blacksquare} \mathcolor{lightgray}{\blacksquare} \mathcolor{lightgray}{\blacksquare} \blacksquare $  &   \multirow{2}{*}{0.77}& \multirow{2}{*}{0.70} & \multirow{2}{*}{0.65} &  \multirow{2}{*}{0.56}  & \multirow{2}{*}{0.79} & \multirow{2}{*}{0.94} & \multirow{2}{*}{0.65} & \cellcolor{LightRed} \\
{\bf gw/mw windows} & {\tt mw}: $\blacksquare  \blacksquare \mathcolor{lightgray}{\blacksquare} \mathcolor{lightgray}{\blacksquare} \mathcolor{lightgray}{\blacksquare}  \mathcolor{lightgray}{\blacksquare}$  & {\tt mw}: $\mathcolor{lightgray}{\blacksquare} \mathcolor{lightgray}{\blacksquare} \blacksquare \blacksquare \mathcolor{lightgray}{\blacksquare} \mathcolor{lightgray}{\blacksquare} $  &  & &  & & & & & \cellcolor{LightRed}  \multirow{-2}{*}{\bf 0.93}\\
            \hline

\multirow{1}{*}{\bf Temporally consistent}& {\tt gw}: $\blacksquare \blacksquare \mathcolor{lightgray}{\blacksquare} \mathcolor{lightgray}{\blacksquare} \mathcolor{lightgray}{\blacksquare}  \mathcolor{lightgray}{\blacksquare}$  & {\tt gw}: $\mathcolor{lightgray}{\blacksquare} \mathcolor{lightgray}{\blacksquare} \blacksquare \blacksquare \blacksquare \blacksquare $
& \cellcolor{LightGreen}  & \cellcolor{LightGreen} & \cellcolor{LightGreen}  & \cellcolor{LightGreen} & \multirow{2}{*}{0.62} & \multirow{2}{*}{0.94} & \multirow{2}{*}{0.33} & \multirow{2}{*}{0.96} \\
                				  ({\it realistic}) & {\tt mw}: $\blacksquare \blacksquare \mathcolor{lightgray}{\blacksquare} \mathcolor{lightgray}{\blacksquare} \mathcolor{lightgray}{\blacksquare}  \mathcolor{lightgray}{\blacksquare}$  & {\tt mw}: $\mathcolor{lightgray}{\blacksquare} \mathcolor{lightgray}{\blacksquare} \blacksquare \blacksquare \blacksquare \blacksquare $  & \multirow{-2}{*}{\cellcolor{LightGreen} {\color{black} \bf 0.58}} & \multirow{-2}{*}{\cellcolor{LightGreen} { \color{black} \bf 0.45}}  &  \multirow{-2}{*}{\cellcolor{LightGreen}{\color{black}\bf 0.32}}   &  \multirow{-2}{*}{\cellcolor{LightGreen} {  \color{black} \bf 0.30}} & & & & \\
            \hline

        \end{tabular}%
        }

        \caption{$F_1$-Score results that show impact of spatial (in columns) and temporal (in rows) experimental bias. Values with red backgrounds are experimental results of (unrealistic) settings similar to those considered in papers of \drebin{}~\cite{Arp:Drebin} and \mmd{}~\cite{Mariconti:MaMaDroid}; values with green background (last row) are results in the realistic settings we identify. The dataset consists of three years (\autoref{sec:dataset}), and each square on the left part of the table represents a six month time-frame: if training (resp. testing) objects are sampled from that time frame, we use a black square ($\blacksquare$); if not, we use a gray square ($\mathcolor{lightgray}{\blacksquare}$).}
        \label{tab:bias}

    \end{table*}

\section{Sources of Experimental Bias}
\label{sec:bias}

In this section, we motivate our discussion of bias through
experimentation with \drebin{}~\cite{Arp:Drebin} and \mmd{}~\cite{Mariconti:MaMaDroid} (\autoref{sec:problem}). We then
detail the sources of temporal (\autoref{sec:bias-time}) and spatial bias
(\autoref{sec:bias-space}) that affect ML-based Android malware classification.

\subsection{Motivational Example}
\label{sec:problem}

We consider a motivational example in which we vary the sources of experimental
bias to better illustrate the problem. \autoref{tab:bias} reports the
$F_1$-score for \drebin{} and \mmd{} under various experimental configurations;
rows correspond to different sources of temporal experimental bias, and columns
correspond to different sources of spatial experimental bias. On the left-part of
\autoref{tab:bias}, we use squares
($\blacksquare/\mathcolor{lightgray}{\blacksquare}$) to show from which time
frame training and testing objects are taken; each square represents six months
(in the window from Jan 2014 to Dec 2016). Black squares ($\blacksquare$) denote
that samples are taken from that six-month time frame, whereas periods with gray
squares ($\mathcolor{lightgray}{\blacksquare}$) are not used. The columns on the
right part of the table correspond to different percentages of malware in the
training set~$Tr$ and the testing set~$Ts$.

\autoref{tab:bias} shows that both \drebin{} and \mmd{} perform far worse in
realistic settings (bold values with green background in the last row, for columns corresponding to 10\% malware in testing) than in settings similar to those presented
in~\cite{Arp:Drebin,Mariconti:MaMaDroid} (bold values with red background). This
is due to inadvertent experimental bias as outlined in the following.

{\bf Note.}
We clarify to which similar settings of~\cite{Arp:Drebin,Mariconti:MaMaDroid} we refer to in the cells with red background in \autoref{tab:bias}. The paper of \mmd{}~\cite{Mariconti:MaMaDroid} reports in the abstract performance ``up to 99\% $F_1$'', which (out of the many settings they evaluate) corresponds to a scenario with 86\% malware in both training and testing, evaluated with 10-fold CV; here, we rounded off to 90\% malware for a cleaner presentation (we have experimentally verified that results with 86\% and 90\% malware-to-benign class ratio are similar).
 \drebin{}'s original paper~\cite{Arp:Drebin} relies on hold-out by performing 10 random splits (66\% training and 33\% testing). Since hold-out is almost equivalent to k-fold CV and suffers from the same spatio-temporal biases, for the sake of simplicity in this section we refer to a k-fold CV setting for both~\drebin{} and~\mmd{}.

\subsection{Temporal Experimental Bias}
\label{sec:bias-time}

\emph{Concept drift} is a problem that occurs in machine learning when a model becomes obsolete as the distribution of incoming data at test-time differs from that of training data, i.e., when the assumption does not hold that data is independent and identically distributed (i.i.d.)~\cite{Jordaney:Transcend}. In the ML community, this problem is also known as \emph{dataset shift}~\cite{DatasetShift}. \emph{Time decay} is the decrease in model performance over time caused by concept drift.

Concept drift in malware combined with similarities among malware within the
same family causes \emph{k-fold cross validation} (CV) to be \emph{positively
  biased}, artificially inflating the performance of malware
classifiers~\cite{Allix:Timeline,Miller:Reviewer,Miller:Thesis}.  K-fold CV is
likely to include in the training set at least one sample of each malware family
in the dataset, whereas new families will be unknown at training time in a
real-world deployment. The all-black squares in \autoref{tab:bias} for 10-fold
CV refer to each training/testing fold of the 10 iterations containing at least
one sample from each time frame.
The use of k-fold CV is widespread in malware classification
research~\cite{Miller:Thesis,Perdisci:URLs,Suarez:DroidSieve,Maggi:URLs,Dahl:Large,Tahan:Mal,Markel:Building,Laskov:PDF,Zhang:Semantics,Curtsinger:Zozzle};
while a useful mechanism to prevent overfitting~\cite{Bishop:ML} or estimate the performance of a classifier in the \emph{absence} of concept drift when the i.i.d. assumption holds (see considerations in \autoref{sec:theexample}), it
has been unclear how it affects the real-world performance of machine learning
techniques with non-stationary data that are affected by time decay. Here, in the first row of Table~\ref{tab:bias}, we quantify the performance impact in the Android domain.

The second row of \autoref{tab:bias} reports an experiment in which a
classifier's ability to detect past objects is
evaluated~\cite{Allix:Timeline,Mariconti:MaMaDroid}. Although this
characteristic is important, high performance should be expected from a
classifier in such a scenario: if the classifier contains at least one
variant of a past malware family, it will likely identify
similar variants. We thus believe that experiments on the performance
achieved on the detection of past malware can be misleading; the
community should focus on building malware classifiers that are robust
against time decay. 

In the third row, we identify a novel temporal bias that occurs when goodware and malware correspond to different time periods, often due to having originated from different data sources (e.g., in~\cite{Mariconti:MaMaDroid}). The black and gray squares in \autoref{tab:bias} show that, although malware testing objects are posterior to malware training objects, the goodware/malware time windows do not overlap; in this case, the classifier may learn to distinguish applications from different time periods, rather than goodware from malware---again leading to artificially high performance. For instance, spurious features such as new API methods may be able to strongly distinguish objects simply because malicious applications predate that API.

The last row of \autoref{tab:bias} shows that the realistic setting, where
training is temporally precedent to testing, causes the worst classifier
performance in the majority of cases. We present decay plots and a more detailed discussion in~\autoref{sec:rules}.

\subsection{Spatial Experimental Bias}
\label{sec:bias-space}

We identify two main types of spatial experimental bias based on assumptions on
percentages of malware in testing and training sets. All experiments in this
section assume temporal consistency. The model is trained on 2014 and tested on
2015 and 2016 (last row of \autoref{tab:bias}) to allow the analysis
of spatial bias without the interference of temporal bias.

\begin{figure*}[t]
\centering
	\subfigure[\drebin{}: Train with 10\% mw]{
		\includegraphics[width=0.23\textwidth]{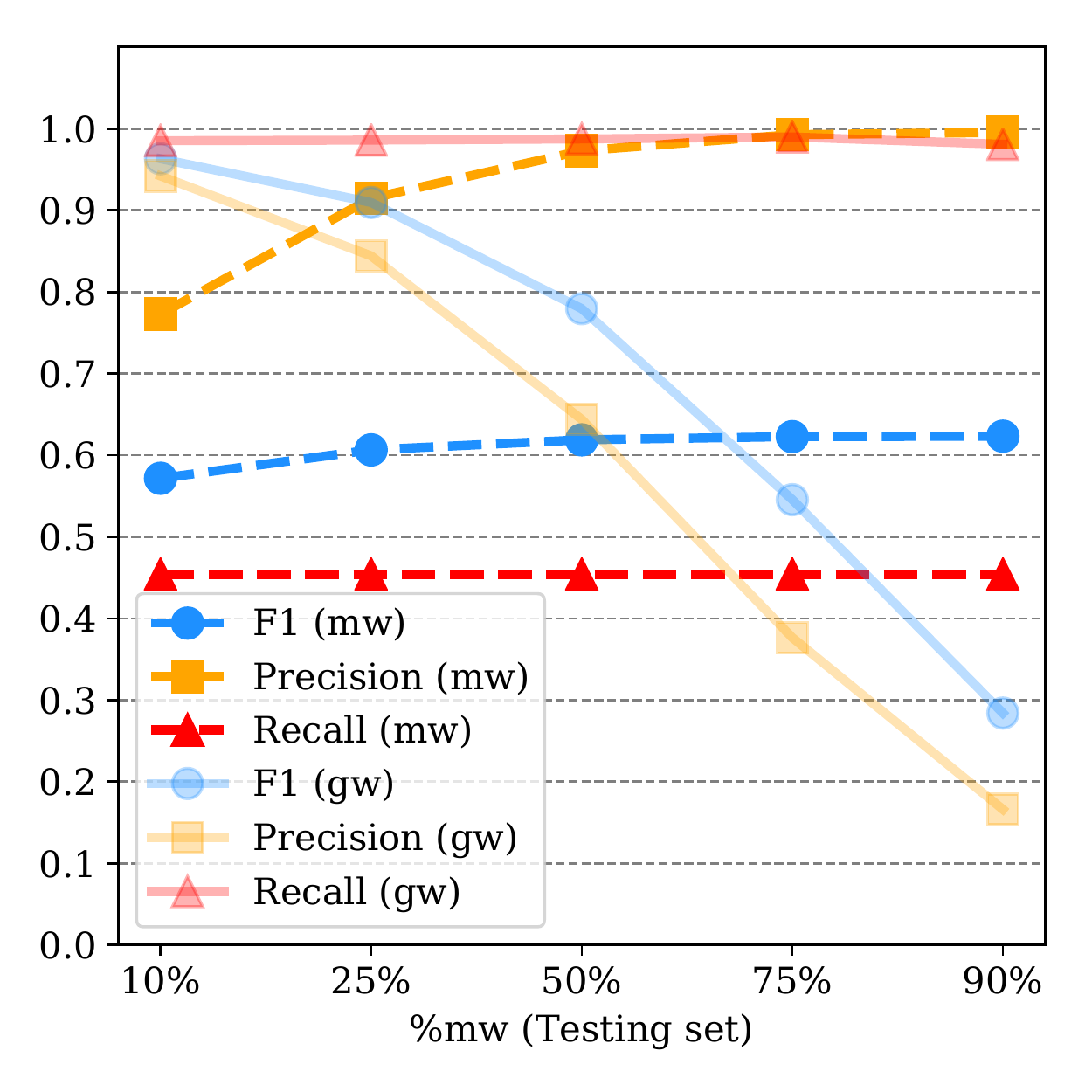}
	}
	\subfigure[\drebin{}: Train with 90\% mw]{
		\includegraphics[width=0.23\textwidth]{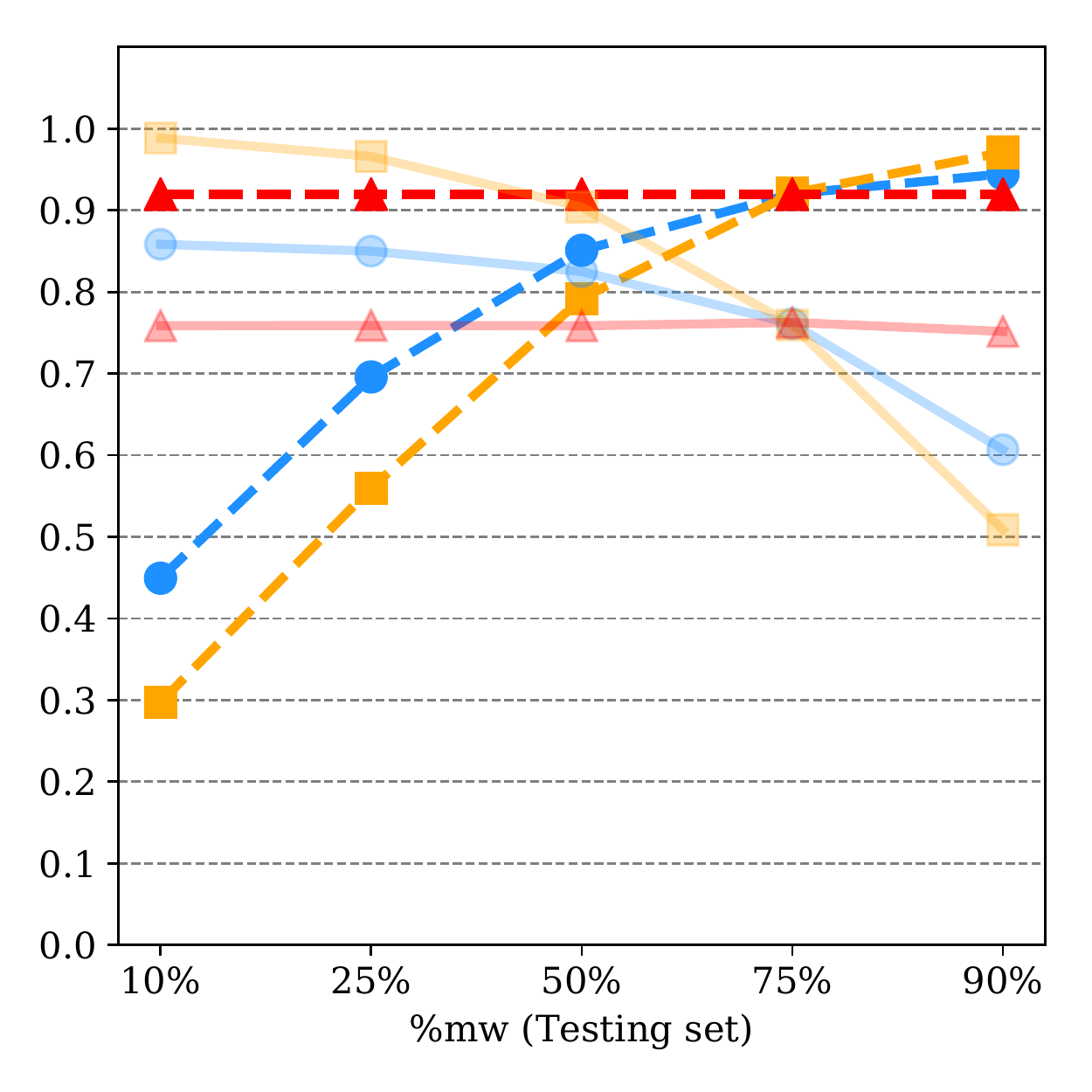}
	}
	\subfigure[\mmd{}: Train with 10\% mw]{
		\includegraphics[width=0.23\textwidth]{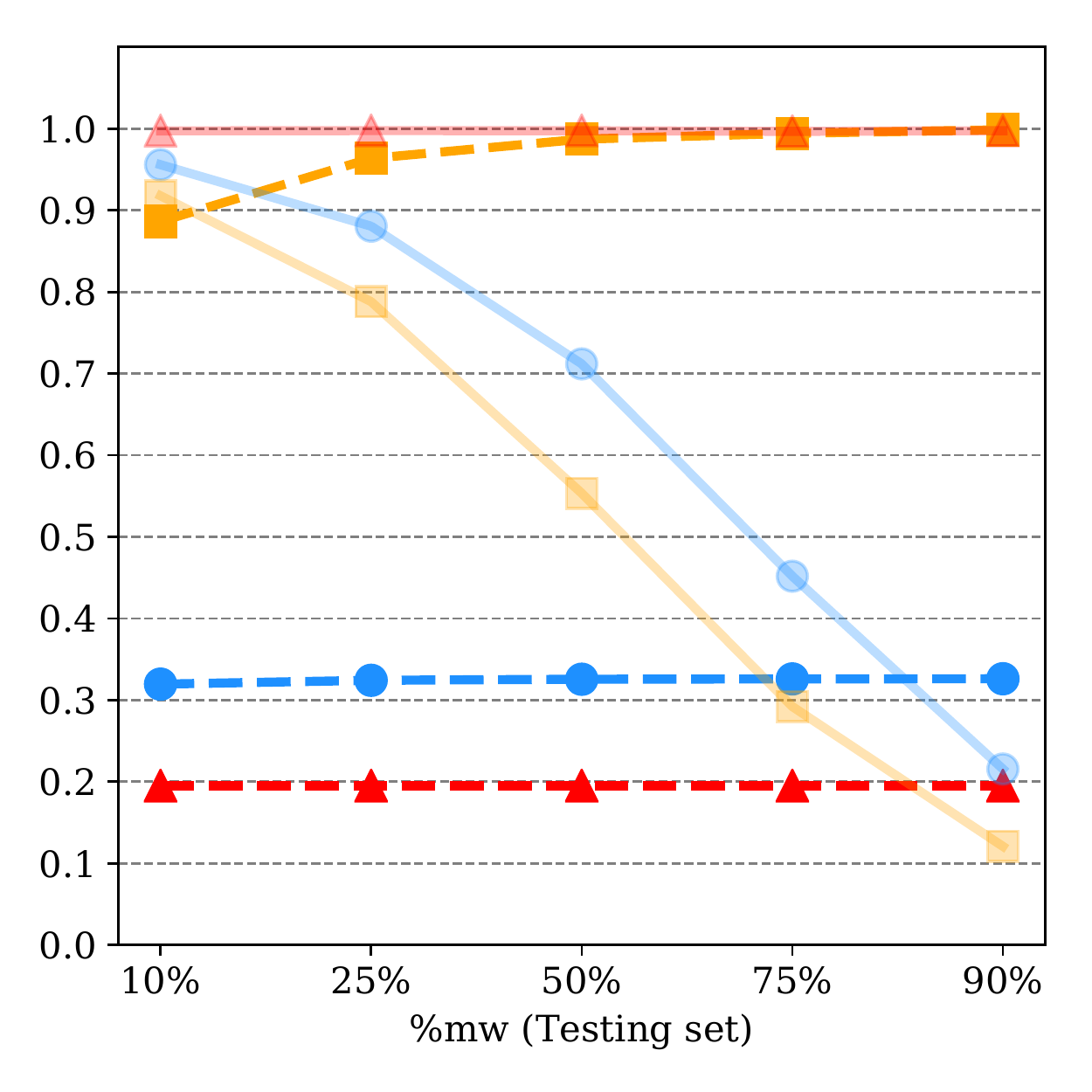}
	}
	\subfigure[\mmd{}: Train with 90\% mw]{
		\includegraphics[width=0.23\textwidth]{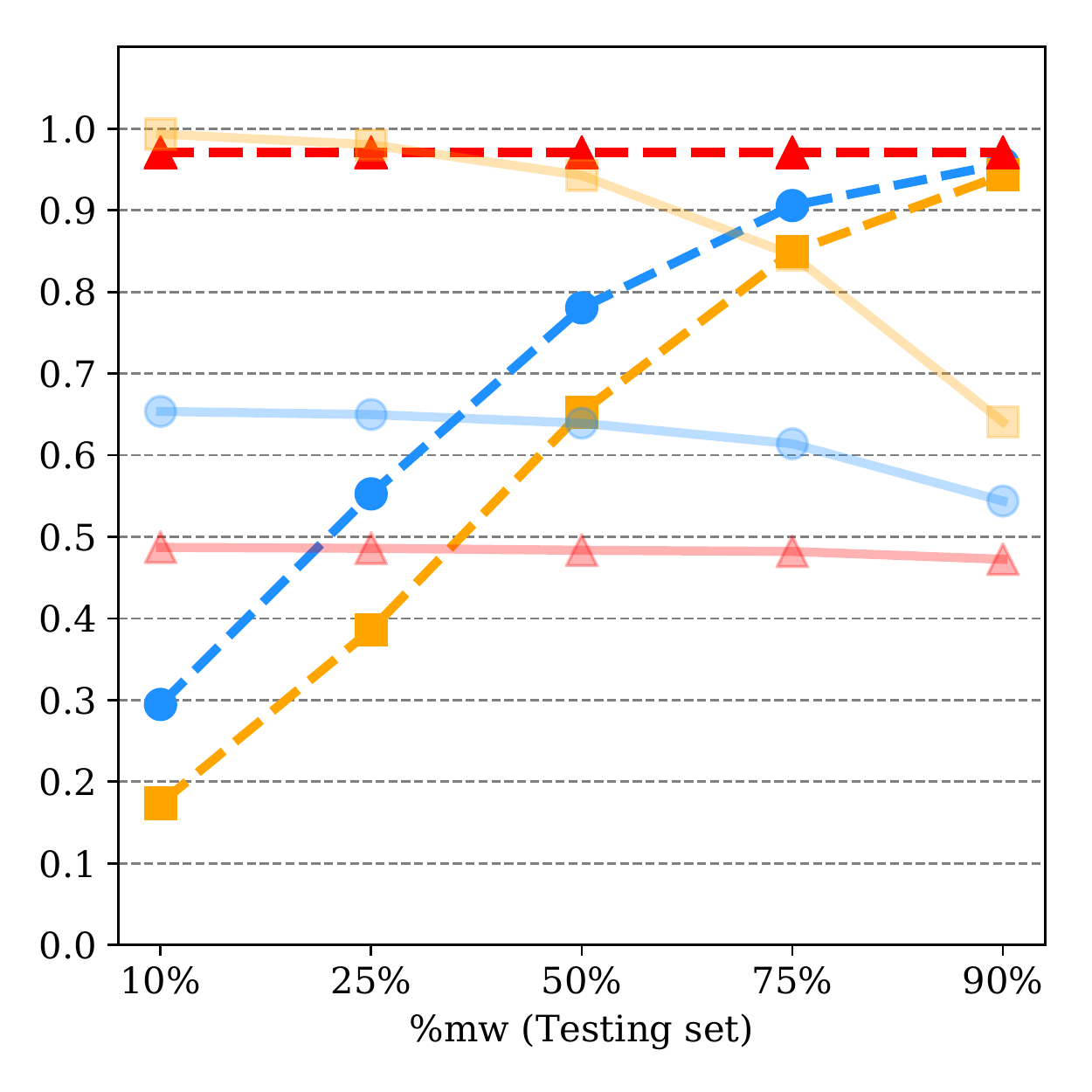}
	}
	\vspace{-10px}
	\caption{{\it Spatial experimental bias in testing}. Training on 2014 and          testing on 2015 and 2016. For increasing \% of malware in the testing (unrealistic setting), Precision for malware increases and Recall remains the same; overall, $F_1$-Score increases for increasing percentage of malware in the testing. However, having more malware than goodware in testing does not reflect the in-the-wild distribution of 10\% malware (\autoref{sec:tsdist}), so the setting with more malware is unrealistic and reports biased results.}
	\label{fig:space-test-bias}
\end{figure*}

{\bf Spatial experimental bias in testing.} The percentage of malware in the testing
distribution needs to be estimated (\autoref{sec:tsdist}) and \emph{cannot} be
changed, if one wants results to be representative of in-the-wild deployment of
the malware classifier. To understand why this leads to biased results, we artificially
vary the testing distribution to illustrate our
point. \autoref{fig:space-test-bias} reports performance ($F_1$-Score,
Precision, Recall) for increasing the percentage of malware during testing on
the $X$-axis. We change the percentage of malware in the testing set by randomly
downsampling goodware, so the number of malware remains fixed throughout the
experiments.\footnotemark[2] For completeness, we report the two training
settings from \autoref{tab:bias} with $10\%$ and $90\%$ malware,
respectively.

Let us first focus on the malware performance (dashed lines).
All plots in \autoref{fig:space-test-bias} exhibit constant Recall, and
increasing Precision for increasing percentage of malware in the
testing. Precision for the malware (mw) class---the positive class---is defined
as $P_{mw}=\sfrac{TP}{(TP+FP)}$ and Recall as $R_{mw}=\sfrac{TP}{(TP+FN)}$. In
this scenario, we can observe that TPs (i.e., malware objects correctly
classified as malware) and FNs (i.e., malware objects incorrectly classified as
goodware) do not change, because the number of malware does not increase; hence,
Recall remains stable. The increase in number of FPs (i.e., goodware objects
misclassified as malware) decreases as we reduce the number of goodware in the
dataset; hence, Precision improves. Since the $F_1$-Score is the harmonic mean
of Precision and Recall, it goes up with Precision. We also observe that, inversely, the Precision for the goodware (gw) class---the negative
class---$P_{gw}=\sfrac{TN}{(TN+FN)}$ decreases (see yellow solid lines in
\autoref{fig:space-test-bias}), because we are reducing the TNs while the FNs do
not change. This example shows how considering an unrealistic testing
distribution with more malware than goodware in this context (\autoref{sec:tsdist}) positively
inflates Precision and hence the $F_1$-Score of the malware
classifier. 

\footnotetext[2]{We choose to downsample goodware to achieve up to
  90\% of malware (mw) for testing because of the computational and storage
  resources required to achieve such a ratio by oversampling.
  This does not alter the
conclusions of our analysis. Let us assume a scenario in which we keep
the same number of goodware (gw), and increase the percentage of mw in the
dataset by oversampling mw. The precision ($P_{mw}=\sfrac{TP}{(TP+FP)}$)
would increase because TPs would increase for any mw detection, and
FPs would not change---because the number of gw remains the same; if
training (resp. testing) observations are sampled from a
distribution similar to the mw in the original dataset (e.g., new training mw
is from 2014 and new testing mw comes from 2015 and 2016), then Recall
($R_{mw}=\sfrac{TP}{(TP+FN)}$) would be stable---it would have the same
proportions of TPs and FNs because the classifier will have a similar
predictive capability for finding mw. Hence, if the number of
mw in the dataset increases, the $F_1$-Score would increase as
well, because Precision increases while Recall remains stable.}\addtocounter{footnote}{1}

\begin{figure*}[t]
\centering
	\subfigure[Training 10\% mw; Testing 10\% mw]{
		\includegraphics[width=0.3\textwidth]{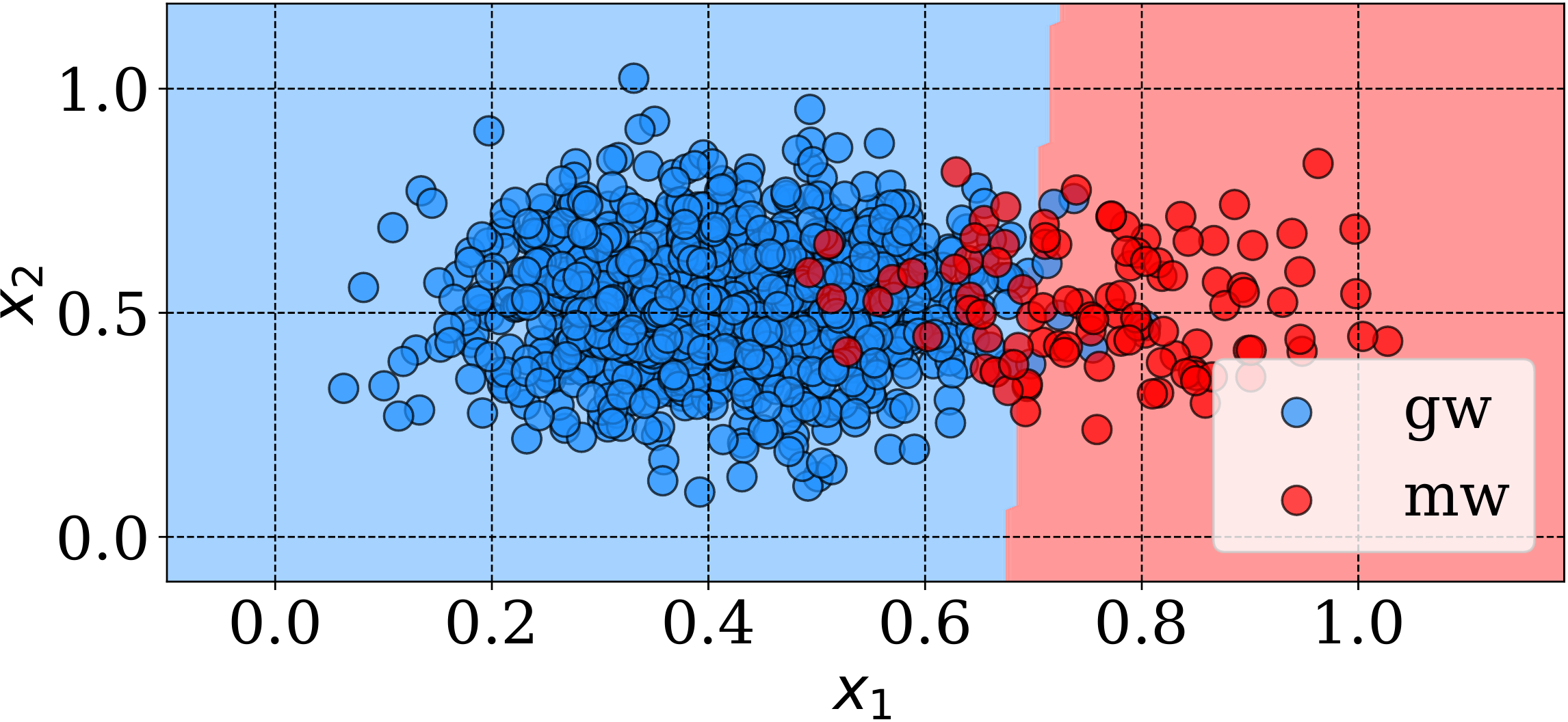}
	}
	\subfigure[Training 50\% mw; Testing 10\% mw]{
		\includegraphics[width=0.3\textwidth]{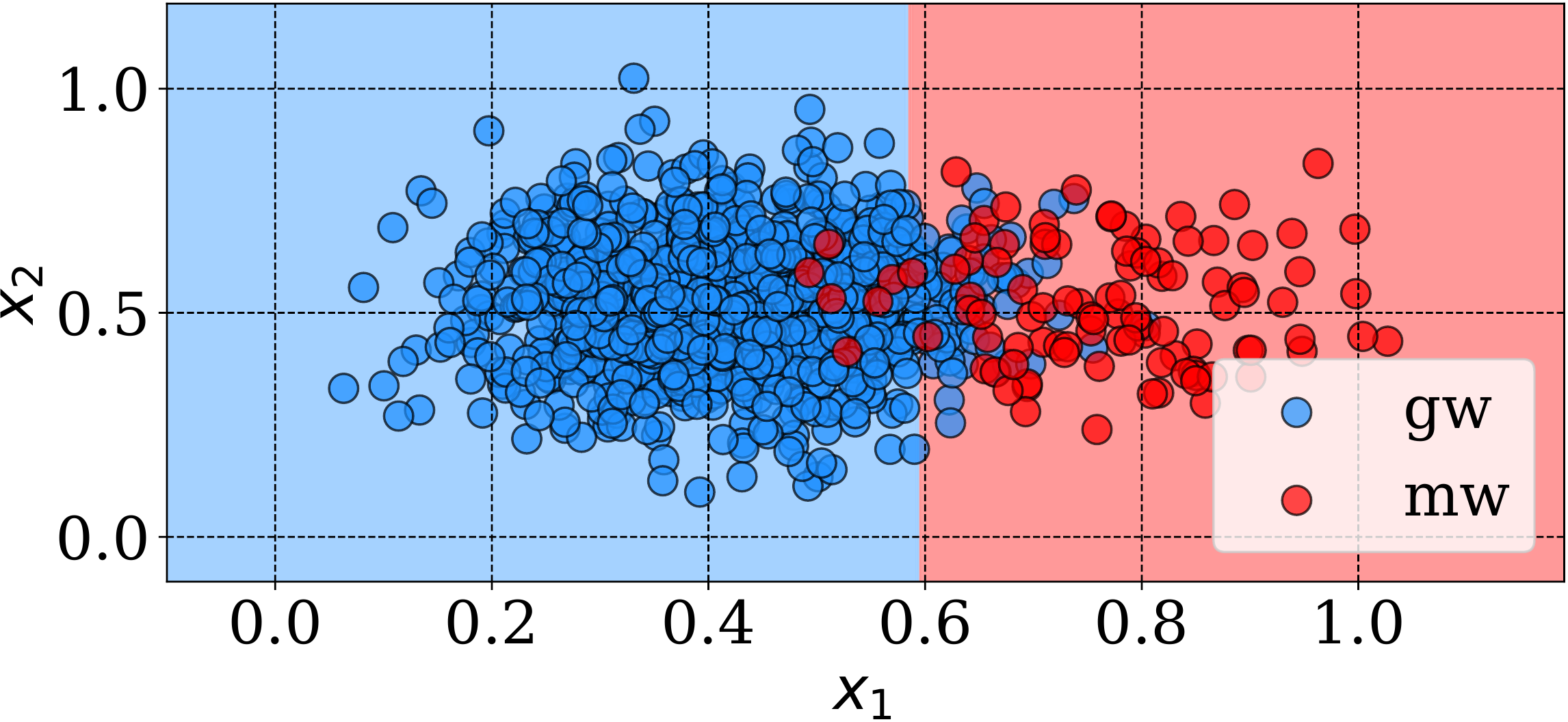}
	}
	\subfigure[Training 90\% mw; Testing 10\% mw]{
	\includegraphics[width=0.3\textwidth]{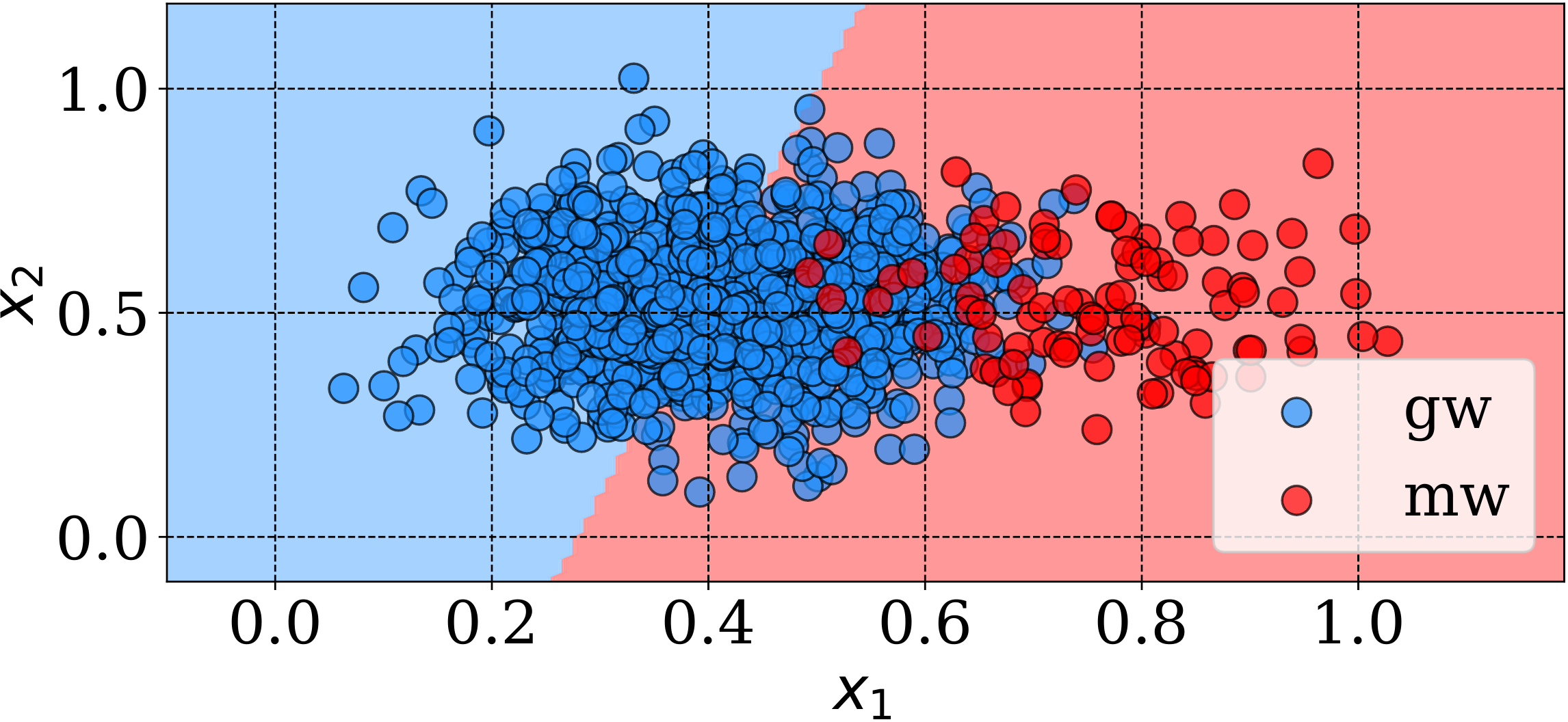}
	}
	\vspace{-10px}
	\caption{Motivating example for the intuition of \emph{spatial experimental bias in training} with Linear-SVM and two features, $x_1$ and $x_2$. The training changes, but the testing points are fixed: 90\% gw and 10\% mw. When the \% of malware in the training increases, the decision boundary moves towards the goodware class, improving Recall for malware but decreasing Precision.}
	\label{fig:toysvm}
\end{figure*}

{\bf Spatial experimental bias in training.}
To understand the impact of altering
malware-to-goodware ratios in training, we now consider a motivating
example with a linear SVM in a 2D feature space, with features $x_1$
and $x_2$. Figure~\ref{fig:toysvm} reports three scenarios, all with
the same $10\%$ malware in testing, but with $10\%$,
$50\%$, and $90\%$ malware in~training.

We can observe that with an increasing
percentage of malware in training, the hyperplane moves towards goodware. More formally, it improves Recall of malware while
reducing its Precision. The opposite is true for goodware. To minimize
the overall error rate $Err=\sfrac{(FP+FN)}{(TP+TN+FP+FN)}$ (i.e.,
maximize Accuracy), one should train the dataset with the same
distribution that is expected in the testing. However, in this
scenario one may have more interest in finding objects of the
minority class (e.g., ``more malware'') by improving Recall subject to
a constraint on maximum FPR.

\begin{figure*}[t]
	\centering
	\subfigure[\drebin{}: Test with 10\% mw]{
		\includegraphics[width=0.23\textwidth]{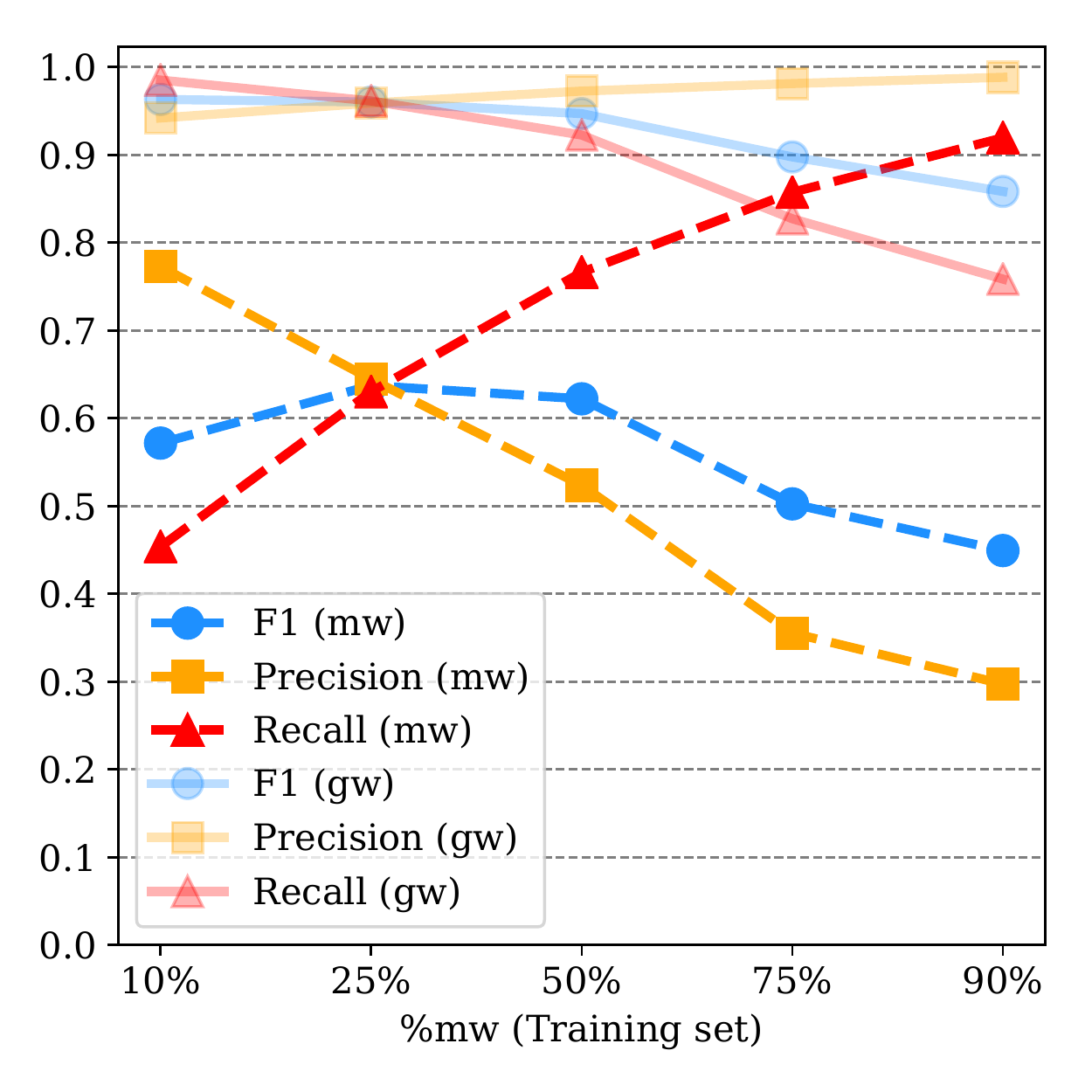}
	}
	\subfigure[\drebin{}: Test with 90\% mw]{
		\includegraphics[width=0.23\textwidth]{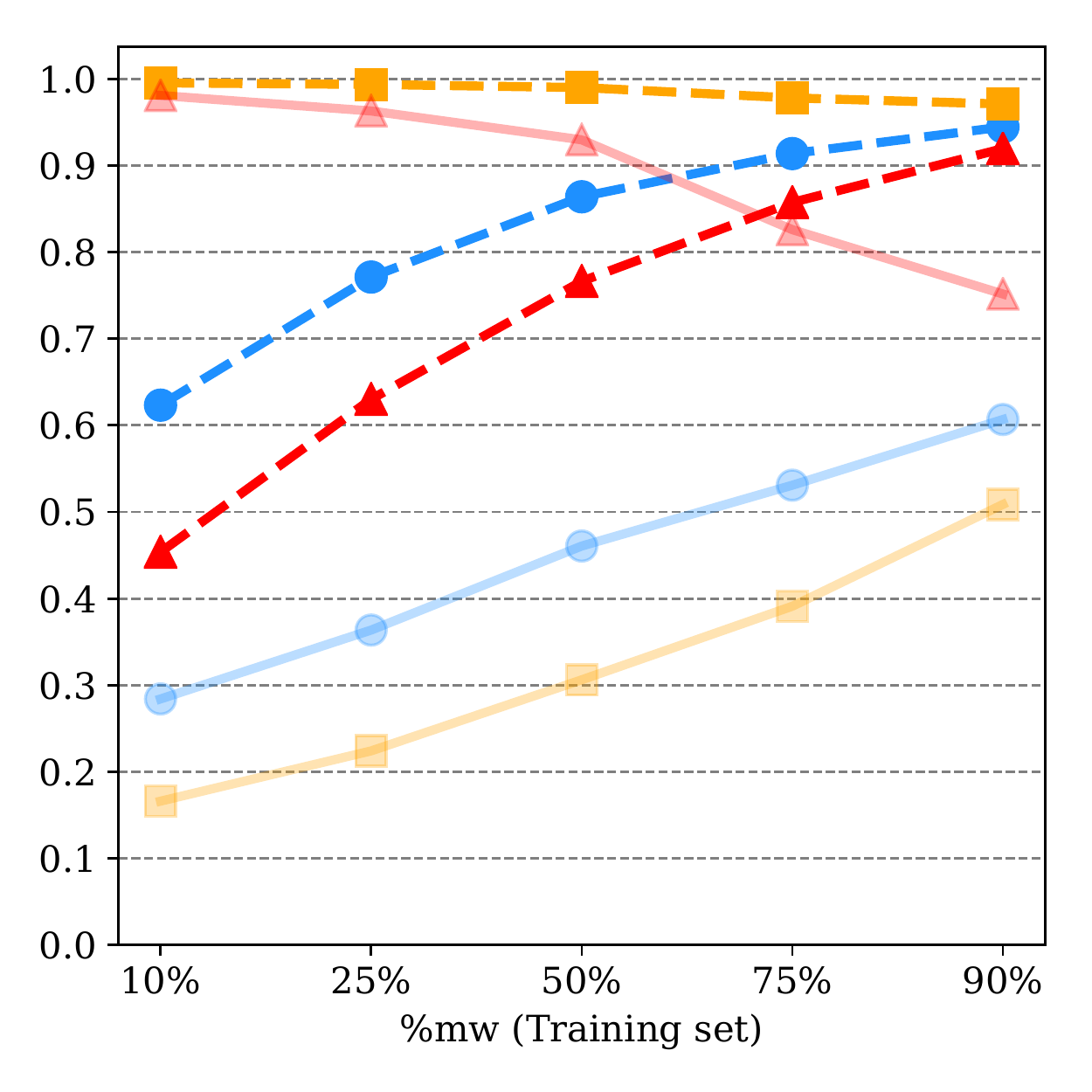}
	}
	\subfigure[\mmd{}: Test with 10\% mw]{
		\includegraphics[width=0.23\textwidth]{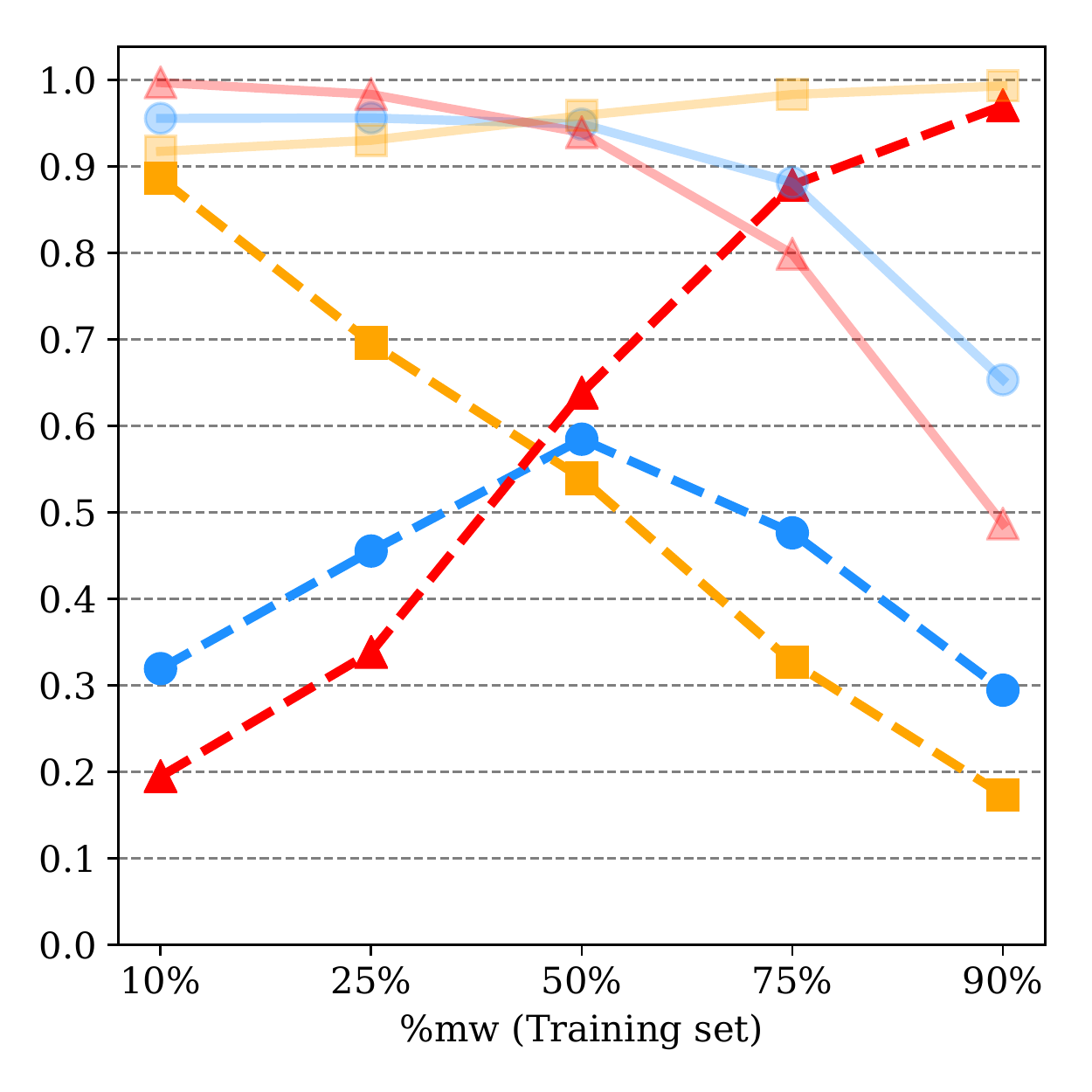}
	}
	\subfigure[\mmd{}: Test with 90\% mw]{
		\includegraphics[width=0.23\textwidth]{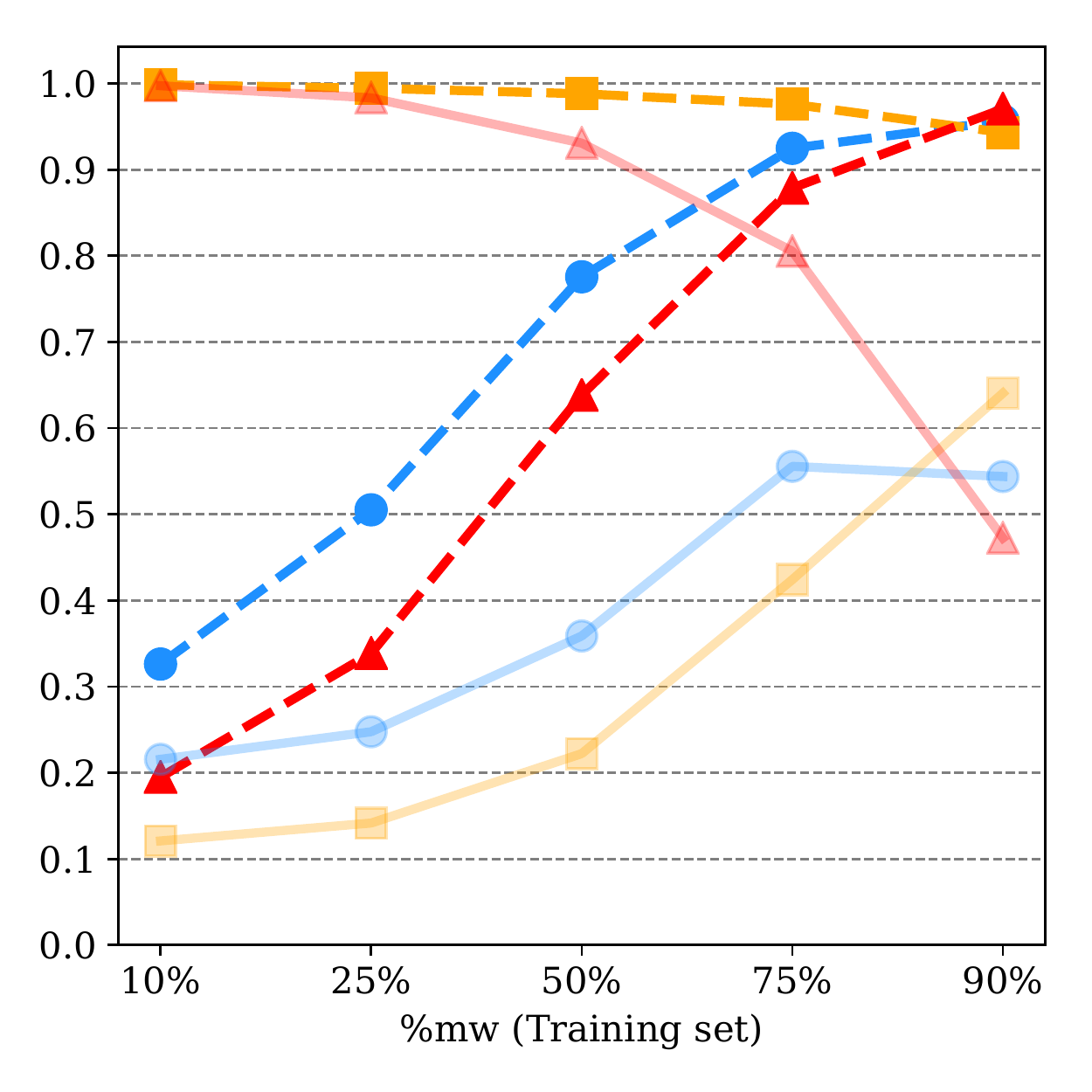}
	}
	\vspace{-10px}
	\caption{{\it Spatial experimental bias in training.} Training on 2014 and testing on 2015 and 2016. For increasing \% of malware in the training, Precision decreases and Recall increases, for the motivations illustrated in the example of \autoref{fig:toysvm}. In~\autoref{sec:tuning}, we devise an algorithm to find the best training configuration for optimizing Precision, Recall, or $F_1$ (depending on user needs).}
	\label{fig:space-train-bias}
\end{figure*}

\autoref{fig:space-train-bias} shows the
performance for \drebin{} and \mmd{}, for increasing percentages of
malware in training on the $X$-axis; just for completeness (since
one cannot artificially change the test distribution to achieve
realistic evaluations), we report results both for 10\% mw in testing
and for 90\% malware in testing, but we remark that in the Android
setting we have estimated 10\% mw in the wild (\autoref{sec:tsdist}). These plots
confirm the trend in our motivating example (\autoref{fig:toysvm}),
that is, $R_{mw}$ increases but $P_{mw}$ decreases. For the plots with
10\% mw in testing, we observe there is a point in which
$F_1$-Score$_{mw}$ is maximum while the error for the gw class
is within~5\%.

In~\autoref{sec:tuning}, we propose a novel algorithm to
improve the performance of the malware class according to the
objective of the user (high Precision, Recall or $F_1$-Score), subject
to a maximum tolerated error. Moreover, in~\autoref{sec:rules} we
introduce constraints and metrics to guarantee bias-free evaluations,
while revealing counter-intuitive results.

\section{Space-Time Aware Evaluation}
\label{sec:rules}

We now formalize how to perform an evaluation of an Android malware classifier
free from spatio-temporal bias. We define a novel set of constraints that
must be followed for realistic evaluations (\autoref{sec:constraints});
we introduce a novel time-aware metric, AUT, that captures in
one number the impact of time decay on a classifier (\autoref{sec:aut});
we propose a novel tuning algorithm that empirically
optimizes a classifier performance, subject to a maximum tolerated
error (\autoref{sec:tuning}); finally, we introduce \tool{} and provide counter-intuitive results through unbiased evaluations (\autoref{sec:theexample}).
To improve readability, we report in Appendix~\ref{app:symbol} a table with all the major symbols used in the remainder of this paper.

\subsection{Evaluation Constraints}
\label{sec:constraints}

Let us consider \dataset{} as a labeled dataset with two classes: malware (positive class) and goodware (negative class). Let us define $s_i \in $ \dataset{} as an \emph{object} (e.g., Android app) with timestamp $\mathit{time}(s_i)$. To evaluate the classifier, the dataset \dataset{} must be split into a training dataset $\mathit{Tr}$ with a time window of size $W$, and a testing dataset $\mathit{Ts}$ with a time window of size $S$.
Here, we consider $S > W$ in order to estimate long-term performance and robustness to decay of the classifier. A user may consider different time splits depending on his objectives, provided each split has a significant number of samples. We emphasize that, although we have the labels of objects in $\mathit{Ts} \subseteq $ \dataset{}, all the evaluations and tuning algorithms \emph{must} assume that labels $y_i$ of objects $s_i \in \mathit{Ts}$ are unknown.

To evaluate performance over time, the test set $\mathit{Ts}$ must be split into time-slots of size $\Delta$. For example, for a testing set time window of size $S = 2 \:\, \mathrm{years}$, we may have $\Delta = 1 \:\, \mathrm{month}$. This parameter is chosen by the user, but it is important that the chosen granularity allows for a statistically significant number of objects in each test window $[t_i, t_i+ \Delta)$.

We now formalize three constraints that must be enforced when dividing
\dataset{} into $\mathit{Tr}$ and $\mathit{Ts}$ for a realistic
setting that avoids spatio-temporal experimental bias (\autoref{sec:bias}).
While C1 was proposed in past work~\cite{Allix:Timeline,
  Miller:Reviewer},
  we are the first to propose C2 and C3---which we show to be fundamental in~\autoref{sec:theexample}.

{\bf C1) Temporal training consistency.} All the
objects in the training must be {\it strictly} temporally
precedent to the testing~ones:
\begin{equation}
	\mathit{time}(s_i) < \mathit{time}(s_j), \forall s_i \in \mathit{Tr} , \forall s_j \in \mathit{Ts}
	\label{eq:c1}
\end{equation}
where $s_i$ (resp. $s_j$) is an object in the training set $\mathit{Tr}$
(resp. testing set $\mathit{Ts}$). Eq.~\ref{eq:c1} must hold; its
violation inflates the results by including
future knowledge in the classifier (\autoref{sec:bias-time}).

{\bf C2) Temporal gw/mw windows consistency.} In every
testing slot of size $\Delta$, all test objects must be from the same
time window: 
\begin{equation}
	t^{\mathit{min}}_i \leq \mathit{time}(s_k) \leq t^{\mathit{max}}_i,\quad \forall s_k \text{ in time slot } [t_i, t_i+\Delta)
	\label{eq:c2}
\end{equation}
where $t^{\mathit{min}}_i = \min_k \mathit{time}(s_k)$ and
$t^{\mathit{max}}_i = \max_k \mathit{time}(s_k)$. The same should hold for the training: although violating Eq.~\ref{eq:c2} in the training data does not bias the evaluation, it may affect the sensitivity of the classifier to unrelated artifacts.
Eq.~\ref{eq:c2} has been violated in the past when goodware and malware have been collected from different time windows (e.g., \mmd{}~\cite{Mariconti:MaMaDroid}, re-evaluated in~\autoref{sec:theexample})---if violated, the results are biased because the classifier may learn and test on artifactual behaviors that, for example, distinguish goodware from malware just by their different API versions.

{\bf C3) Realistic malware-to-goodware ratio in testing.}
Let us define~$\varphi$ as the average percentage of malware in training data, and $\delta$ as the average percentage of malware in the testing data.  Let $\hat{\sigma}$ be the estimated percentage of malware in the wild. To have a realistic evaluation, the average percentage of malware in the testing ($\delta$) must be as close as possible to the percentage of malware in the wild ($\hat{\sigma}$), so that:
\begin{equation}
	\delta \simeq \hat{\sigma}
	\label{eq:c3}
\end{equation}
For example, we have estimated that in the Android scenario goodware is predominant over malware, with $\hat{\sigma} \approx 0.10$ (\autoref{sec:tsdist}). If C3 is violated by overestimating the percentage of malware, the results are positively inflated (\autoref{sec:bias-space}).
We highlight that, although the testing distribution $\delta$ cannot be changed (in order to get realistic results), the percentage of malware in the training $\varphi$ may be tuned (\autoref{sec:tuning}).

\subsection{Time-aware Performance Metrics}
\label{sec:aut}

We introduce a time-aware performance metric that allows for the comparison of
different classifiers while considering time decay. Let $\Theta$ be a
classifier trained on~$\mathit{Tr}$; we capture the performance of $\Theta$
for each time frame $[t_i, t_i+\Delta)$ of the testing set~$\mathit{Ts}$ (e.g.,
each month). We identify two options to represent per-month performance:
\begin{itemize}[noitemsep,nolistsep]
	\item {\bf Point estimates} (\pnt): The value plotted on the $Y$-axis for $x_k = k \Delta$ (where $k$ is the test slot number) computes the performance metric (e.g., $F_1$-Score) only based on predictions $\hat{y}_i$ of $\Theta$ and true labels $y_i$ in the interval $[W+(k-1) \Delta, W + k \Delta)$.
\item {\bf Cumulative estimates} (\cum): The value plotted on the $Y$-axis for $x_k = k \Delta$ (where $k$ is the test slot number) computes the performance metric (e.g., $F_1$-Score) only based on predictions $\hat{y}_i$ of $\Theta$ and true labels $y_i$ in the cumulative interval $[W, W + k \Delta)$.
\end{itemize}
Point estimates are always to be preferred to represent the real performance of an algorithm. The cumulative estimates can be used to highlight a smoothed trend and to show overall performance up to a certain point, but can be misleading if reported on their own if objects are too sparsely distributed in some test slots $\Delta$. Hence, we report only point estimates in the remainder of the paper (e.g., in \autoref{sec:theexample}), while an example of cumulative estimate plots is reported in Appendix~\ref{sec:app:cml}. 

To facilitate the comparison of different time decay plots, we define a new
metric, {\it Area Under Time} ({\bf AUT}), the area under the performance
curve over time. Formally, based on the trapezoidal rule (as in AUROC~\cite{Bishop:ML}), AUT is defined as~follows:
\begin{equation}
\displaystyle AUT(f,N) = \frac{1}{N-1} \sum_{k=1}^{N-1}  \frac{[f(x_{k+1}) + f(x_k)]}{2} \label{eq:aut}
\end{equation}
where: $f(x_k)$ is the value of the point estimate of the performance metric $f$ (e.g., $F_1$) evaluated at point~$x_k := (W + k \Delta)$; $N$ is the number of test slots, and $1/(N-1)$ is a normalization factor so that AUT $\in [0,1]$.
The perfect classifier with robustness to time decay in the time
window $S$ has $\mathrm{AUT}=1$. By default, AUT is computed as the area under point estimates, as they capture the trend of the classifier over time more closely; if the AUT is computed on cumulative estimates, it should be explicitly marked as $\mathrm{AUT}_{\cum}$. As an example, $\mathrm{AUT}(F_1, 12m)$ is the point estimate of $F_1$-Score considering time decay for a period of 12 months, with a 1-month interval.
We highlight that the simplicity of computing the AUT should be seen
as a benefit rather than a drawback; it is a simple yet effective
metric that captures the performance of a classifier with respect to
time decay, de-facto promoting a fair comparison across different
approaches.

\begin{framed}
{\bf AUT}($f$,$N$) is a metric that allows us to evaluate performance $f$ of a malware classifier against time decay over $N$ time units in realistic experimental settings---obtained by enforcing C1, C2, and C3 (\autoref{sec:constraints}). The next sections leverage AUT for tuning classifiers and comparing different solutions (\autoref{sec:theexample}).
\end{framed}

\subsection{Tuning Training Ratio}
\label{sec:tuning}

We propose a novel algorithm that allows for the adjustment of the
training ratio $\varphi$ when the dataset is imbalanced, in order to
optimize a user-specified performance metric ($F_1$, Precision, or
Recall) on the minority class, subject to a maximum tolerated error, while  aiming to reduce time decay.
The high-level intuition of the impact of changing $\varphi$ is described in~\autoref{sec:bias-space}. We also observe that ML literature has shown ROC curves to be misleading on highly imbalanced datasets~\cite{davis2006relationship,He:Imbalanced}. Choosing different thresholds on ROC curves \emph{shifts} the decision boundary, but (as seen in the motivating example of Figure~\ref{fig:toysvm}) re-training with different ratios $\varphi$ (as in our algorithm) also changes the \emph{shape} of the decision boundary, better representing the minority class.

Our tuning algorithm is inspired by one proposed by Weiss and Provost~\cite{Weiss:ReBalance}; they propose a progressive sampling of training objects to collect a dataset that improves AUROC performance of the minority class in an imbalanced dataset. However, they did not take temporal constraints into account (\autoref{sec:bias-time}), and heuristically optimize only AUROC.
Conversely, we enforce C1, C2, C3 (\autoref{sec:constraints}), and rely on AUT to achieve three possible targets for the malware class: higher $F_1$-Score, higher Precision, or higher Recall. Also, we assume that the user already has a training dataset $\mathit{Tr}$ and wants to use as many objects from it as possible, while still achieving a good performance trade-off; for this purpose, we perform a \emph{progressive subsampling} of the goodware class.

Algorithm~\ref{alg:varphi} formally presents our methodology for tuning the parameter $\varphi$ to find the value $\varphi^*_{\mathbb{P}}$ that optimizes $\mathbb{P}$ subject to a maximum error rate \errorratemax{}. The algorithm aims to solve the following optimization problem:
\begin{equation}
    \text{maximize}_\varphi  \{ \mathbb{P} \} \quad \text{subject to: }  E \leq E_{max} \label{eq:optimization}
\end{equation}
where $\mathbb{P}$ is the target performance: the $F_1$-Score ($F_1$), Precision ($Pr$) or Recall ($Rec$) of the malware class; \errorratemax{} is the maximum tolerated error; depending on the target $\mathbb{P}$, the error rate \errorrate{} has a different formulation:
\begin{itemize}[noitemsep, nolistsep]
    \item if $\mathbb{P}=F_1 \rightarrow$ \errorrate{}$=1-\text{Acc}=\sfrac{(FP+FN)}{(TP+TN+FP+FN)}$
    \item if $\mathbb{P}=Rec \rightarrow$ \errorrate{}$=FPR=\sfrac{FP}{(TN+FP)}$
    \item if $\mathbb{P}=Pr \rightarrow$ \errorrate{}$=FNR=\sfrac{FN}{(TP+FN)}$
\end{itemize}
Each of these definitions of \errorrate{} is targeted to limit the error induced by the specific performance---if we want to maximize $F_1$ for the malware class, we need to limit both FPs and FNs; if $\mathbb{P}=Pr$, we increase FNs, so we constrain FNR. 

Algorithm~\ref{alg:varphi} consists of two phases: \emph{initialization} (lines~\ref{alg:varphi:init-start}--\ref{alg:varphi:init-end}) and \emph{grid search} of $\varphi^*_{\mathbb{P}}$ (lines \ref{alg:varphi:core-start}--\ref{alg:varphi:core-end}). In the initialization phase, the training set $Tr$ is split into a proper training set \emph{ProperTr} and a validation set \emph{Val}; this is split according to the space-time evaluation constraints in~\autoref{sec:constraints}, so that all the objects in \emph{ProperTr} are temporally anterior to \emph{Val}, and the malware percentage $\delta$ in \emph{Val} is equal to $\hat{\sigma}$, the in-the-wild malware percentage. The maximum performance observed \performancestar{} and the optimal training ratio $\varphi^*_{\mathbb{P}}$ are initialized by assuming the estimated in-the-wild malware ratio $\hat{\sigma}$ for training; in Android, $\hat{\sigma} \approx 10\%$ (see \autoref{sec:tsdist}).

The grid-search phase iterates over different values of $\varphi$, with a learning rate $\mu$ (e.g., $\mu=0.05$), and keeps as $\varphi^*_{\mathbb{P}}$ the value leading to the best performance, subject to the error constraint. To reduce the chance of discarding high-quality points while downsampling goodware, we prioritize the most uncertain points (e.g., points close to the decision boundary in an SVM)~\cite{Settles:AL}. The constraint on line~\ref{alg:varphi:core-start} ($\hat{\sigma} \leq \varphi \leq 0.5$) is to ensure that one does not under-represent the minority class (if $\varphi < \hat{\sigma}$) and that one does not let it become the majority class (if $\varphi > 0.5$); also, from \autoref{sec:bias-space} it is clear that if $\varphi > 0.5$, then the error rate becomes too high for the goodware class. Finally, the grid-search explores multiple values of $\varphi$ and stores the best ones. To capture time-aware performance, we rely on AUT (\autoref{sec:aut}), and the error rate is computed according to the target $\mathbb{P}$ (see above). Tuning examples are in~\autoref{sec:theexample}.

\begin{algorithm2e}[t]
\caption{Tuning $\varphi$.}
\footnotesize
\label{alg:varphi}
\DontPrintSemicolon
\SetAlgoNoEnd
\KwIn{Training dataset $Tr$}
\SetKwInOut{Parameter}{Parameters}
\Parameter{Learning rate $\mu$, target performance $\mathbb{P} \in \{F_1, Pr, Rec\}$, max error rate \errorratemax{}}
\KwOut{$\varphi^{*}_{\mathbb{P}}$, optimal percentage of mw to use in training to achieve the best target performance $\mathbb{P}$ subject to \errorrate{}$<$\errorratemax{}.}
\hrule
Split the training set \emph{Tr} into two subsets: actual training (\emph{ProperTr}) and validation set (\emph{Val}), while enforcing C1, C2, C3 (\autoref{sec:constraints}), also implying $\delta=\hat{\sigma}$\label{alg:varphi:init-start}\\
Divide \emph{Val} into $N$ non-overlapped subsets, each corresponding to a time-slot $\Delta$, so that \emph{Val}$_{array}$ = $[V_0, V_1,...,V_N]$ \\
Train a classifier $\Theta$ on \emph{ProperTr}\\
\performancestar{} $\gets$ AUT($\mathbb{P}$,$N$) on \emph{Val}$_{array}$ with $\Theta$\\
$\varphi^*_{\mathbb{P}} = \hat{\sigma}$ \label{alg:varphi:init-end}\\
\For{($\varphi=\hat{\sigma}$; $\varphi \leq 0.5$; $\varphi = \varphi + \mu$)}{
\label{alg:varphi:core-start}

Downsample gw in \emph{ProperTr} so that percentage of mw is $\varphi$\\
Train the classifier $\Theta_\varphi$ on \emph{ProperTr} with $\varphi$ mw \\
performance \performance{}$_\varphi$ $\gets$ AUT($\mathbb{P}$, $N$)  on \emph{Val}$_{array}$ with $\Theta_\varphi$\\
error \errorrate{}$_\varphi$ $\gets$ Error rate on \emph{Val}$_{array}$ with $\Theta_\varphi$\\

\If{(\performance{}$_\varphi > $ \performancestar{}) {\bf and} (\errorrate{}$_\varphi \leq$\errorratemax{})} {
	\performancestar{} $\gets $ \performance$_\varphi$\\
	$\varphi^*_{\mathbb{P}} \gets \varphi$\\
}
}
\Return $\varphi^*_{\mathbb{P}}$; \label{alg:varphi:core-end}
\end{algorithm2e}

\begin{figure*}[t]
	\centering
	\includegraphics[width=0.95\textwidth]{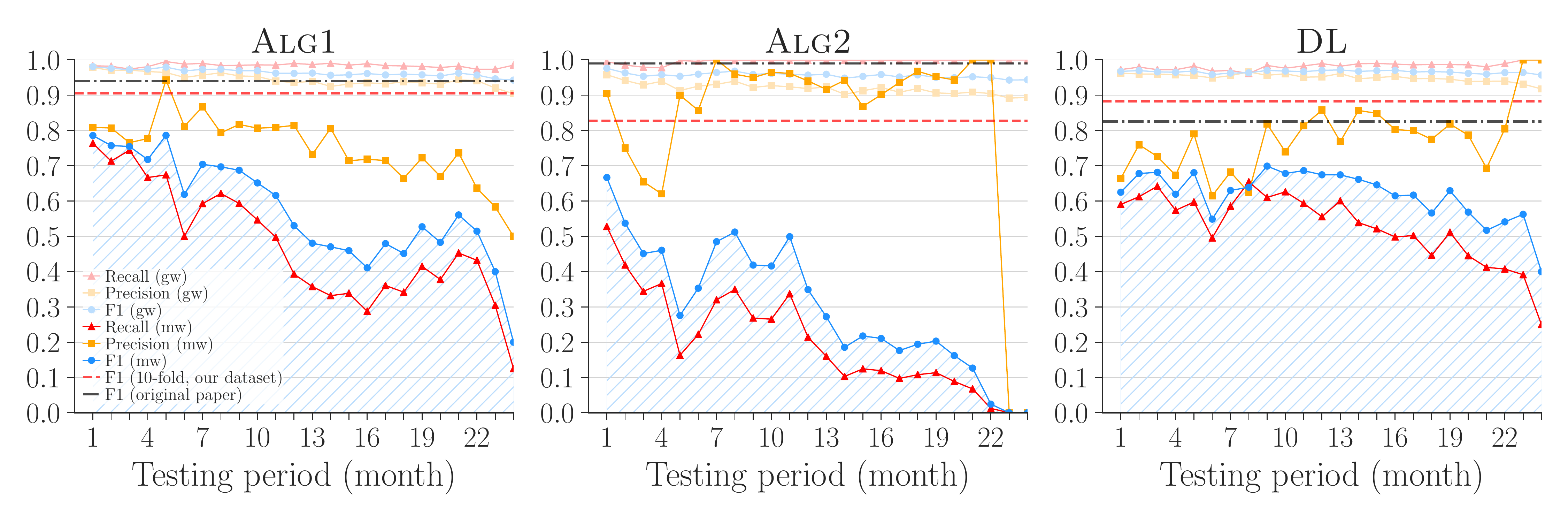}
	\vspace{-15px}
	\caption{Time decay of \drebin{}~\cite{Arp:Drebin}, \mmd{}~\cite{Mariconti:MaMaDroid} and \dl{}~\cite{Papernot:ESORICS}---with AUT$(F_1,24m)$ of 0.58, 0.32 and 0.64, respectively. Training and test distribution both have 10\% malware.
	The drop in the last 3 months is also related to lower samples in the dataset.}
	\label{fig:decay}
\end{figure*}

\subsection{\tool{}: Revealing Hidden Performance}
\label{sec:theexample}

Here, we show how our methodology can reveal hidden performance of \drebin{}~\cite{Arp:Drebin}, \mmd{}~\cite{Mariconti:MaMaDroid}, and \dl{}~\cite{Papernot:ESORICS} (\autoref{sec:algo}), and their robustness to time decay.

We develop \tool{} as an open-source Python framework that enforces constraints C1, C2, and C3 (\autoref{sec:constraints}), computes AUT (\autoref{sec:aut}), and can train a classifier with our tuning algorithm (\autoref{sec:tuning}). \tool{} operates as a traditional Python ML library but, in addition to features matrix~$X$ and labels~$y$, it also takes as input the timestamp array~$t$ containing dates for each object. Details about \tool{}'s implementation and generality are in Appendix~\ref{sec:artifact}.

Figure~\ref{fig:decay} reports several performance metrics of the
three algorithms as point estimates over time. The $X$-axis reports
the testing slots in months, whereas the $Y$-axis reports different
scores between 0 and 1. The areas highlighted in blue correspond to
the AUT($F_1,24m$). The black dash-dotted horizontal lines represent the
best $F_1$ from the original
papers~\cite{Arp:Drebin,Mariconti:MaMaDroid,Papernot:ESORICS},
corresponding to results obtained with 10 hold-out random splits for
\drebin{}, 10-fold CV for \mmd{}, and random split for \dl{}; all
these settings are analogous to k-fold from a temporal bias
perspective, and violate both C1 and C2. The red dashed horizontal lines correspond to 10-fold $F_1$ obtained on our dataset, which satisfies C3.

{\bf Differences in 10-fold $F_1$.} We discuss and motivate the differences between the horizontal lines representing original papers' best $F_1$ and replicated 10-fold $F_1$.  The 10-fold $F_1$ of \drebin{} is close to the original paper~\cite{Arp:Drebin}; the difference is likely related to the use of a different, more recent dataset. The 10-fold $F_1$ of \mmd{} is much lower than the one in the paper. We verified that this is mostly caused by {\bf violating C3}: the best $F_1$ reported in~\cite{Mariconti:MaMaDroid} is on a setting with 86\% malware---hence, spatial bias increases even 10-fold $F_1$ of \mmd{}. Also {\bf violating C2} tends to inflate the 10-fold performance as the classifier may learn artifacts. The 10-fold $F_1$ in \dl{} is instead slightly higher than in the original paper~\cite{Papernot:ESORICS}; this is likely related to a hyperparameter tuning in the original paper that optimized Accuracy (instead of $F_1$), which is known to be misleading in imbalanced datasets. Details on hyperparameters chosen are in~\autoref{sec:hyperparam}. From these results, we can observe that even if an analyst wants to estimate what the performance of the classifier would be in the \emph{absence} of concept drift (i.e., where objects coming from the same distribution of the training dataset are received by the classifier),  she still needs to enforce C2 and C3 while computing 10-fold CV to obtain valid results.

{\bf Violating C1 and C2.} Removing the temporal bias reveals the real
performance of each algorithm in the presence of concept drift. The AUT$(F_1,24m)$ quantifies such performance: 0.58 for \drebin{}, 0.32 for \mmd{} and 0.64 for
\dl{}. In all three scenarios, the AUT$(F_1,24m)$ is lower than
10-fold $F_1$ as the latter violates constraint C1 and may violate C2 if the dataset classes are not evenly distributed across the timeline (\autoref{sec:rules}).

{\bf Best performing algorithm.} \tool{} shows a counter-intuitive result: the algorithm that is most robust to time decay and has the highest performance over the 2 years testing is the \dl{} algorithm (after removing space-time bias), although for the first few months \drebin{} outperforms \dl{}. Given this outcome, one may prefer to use \drebin{} for the first few months and then \dl{}, if retraining is not possible (\autoref{sec:delay}). We observe that this strongly contradicts the performance obtained in the presence of temporal and spatial bias. In particular, if we only looked at the best $F_1$ reported in the original papers, \mmd{} would have been the best algorithm (because spatial bias was present). After enforcing C3, the k-fold on our dataset would have suggested that \dl{} and \drebin{} have similar performance (because of temporal bias). After enforcing C1, C2 and C3, the AUT reveals that \dl{} is actually the algorithm most robust to time decay.

{\bf Different robustness to time decay.}
Given a training dataset, the robustness of different ML models against performance decay over time depends on several factors. Although more in-depth evaluations would be required to understand the theoretical motivations behind the different robustness to time decay of the three algorithms in our setting, we hereby provide insights on possible reasons.
The performance of \mmd{} is the fastest to decay likely because its feature engineering~\cite{Mariconti:MaMaDroid} may be capturing relations in the training data that quickly become obsolete at test time to separate goodware from malware.
Although \drebin{} and \dl{} take as input the same feature space, the higher robustness to time decay of \dl{} is likely related to feature representation in the \emph{latent feature space} automatically identified by deep learning~\cite{Goodfellow:DL}, which appears to be more robust to time decay in this specific setting. Recent results have also shown that linear SVM tends to overemphasize a few important features~\cite{Battista:Explaining}---which are the few most effective on the training data, but may become obsolete over time. We remark that we are \emph{not} claiming that deep learning is always more robust to time decay than traditional ML algorithms. Instead, we demonstrate how, in this specific setting, \tool{} allowed us to highlight higher robustness of \dl{}~\cite{Papernot:ESORICS} against time decay; however, the prices to pay to use \dl{} are lower explainability~\cite{LIME,LEMNA} and higher training time~\cite{Goodfellow:DL}.

\begin{figure}[t]
    \hspace*{-0.3cm}
    \centering
    \includegraphics[width=1.05\columnwidth]{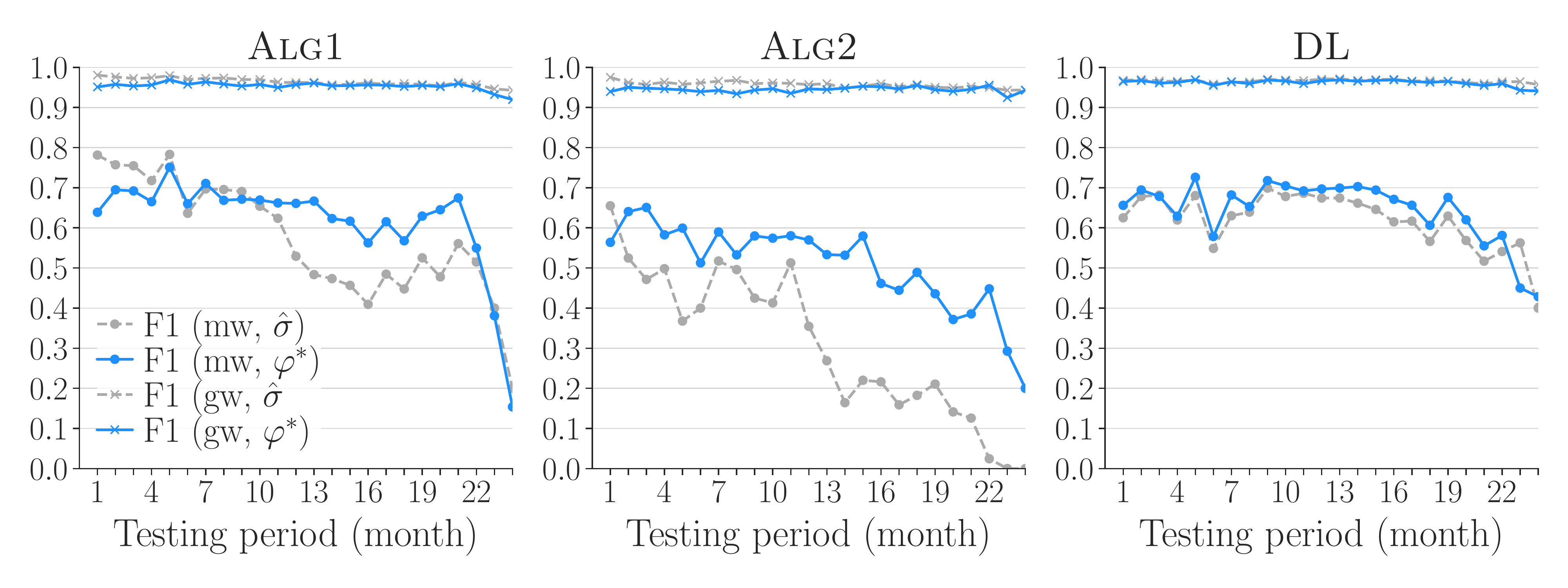}
   	\vspace{-20px}
    \caption{Tuning improvement obtained by applying $\varphi^*_{F_1}=25\%$ to \drebin{} and \dl{}, and $\varphi^*_{F_1}=50\%$ to \mmd{}. The values of $\varphi^*_{F_1}$ are obtained with Algorithm~\ref{alg:varphi} and one year of training data (trained on 8 months and validated on 4 months).}
    \label{fig:varphi}
\end{figure}

{\bf Tuning algorithm.} We now evaluate whether our tuning (Algorithm~\ref{alg:varphi} in \autoref{sec:tuning})   improves robustness to time decay of a malware classifier for a given target performance. We first aim to maximize $\mathbb{P}=F_1$-Score of malware class, subject to \errorratemax{}$=10\%$. After running Algorithm~\ref{alg:varphi} on \drebin{}~\cite{Arp:Drebin}, \mmd{}~\cite{Mariconti:MaMaDroid} and~\dl{}, we find that $\varphi^{*}_{F_1}=0.25$ for \drebin{} and \dl{}, and $\varphi^{*}_{F_1}=0.5$ for \mmd{}. \autoref{fig:varphi} reports the improvement on the test performance of applying $\varphi^{*}_{F_1}$ to the full training set $Tr$ of 1 year. We remark that the choice of $\varphi^{*}_{F_1}$ uses only training information (see Algorithm~\ref{alg:varphi}) and no test information is used---the optimal value is chosen from a 4-month validation set extracted from the 1 year of training data; this is to simulate a realistic deployment setting in which we have no a priori information about testing. \autoref{fig:varphi} shows that our approach for finding the best $\varphi^{*}_{F_1}$ improves the $F_1$-Score on malware at test time, at the cost of slightly reduced goodware performance. \autoref{tab:varphi} shows details of how total FPs, total FNs, and AUT changed by training \drebin{}, \mmd{}, and \dl{} with $\varphi^{*}_{F_1}$, $\varphi^{*}_{Prec}$, and $\varphi^{*}_{Rec}$ instead of $\hat{\sigma}$.
These training ratios have been computed subject to $E_{max} = 5\%$ for $\varphi^{*}_{Rec}$, $E_{max} = 10\%$ for $\varphi^{*}_{F_1}$, and $E_{max} = 15\%$ for $\varphi^{*}_{Prec}$; the difference in the maximum tolerated errors is motivated by the class imbalance in the dataset---which causes lower FPR and higher FNR values (see definitions in \autoref{sec:tuning}), as there are many more goodware than malware.
As expected (\autoref{sec:bias-space}), \autoref{tab:varphi} shows that when training with $\varphi^{*}_{F_1}$ Precision decreases (FPs increase) but Recall increases (because FNs decrease), and the overall AUT increases slightly as a trade-off. A similar reasoning follows for the other performance targets. We observe that the AUT for Precision may slightly differ even with a similar number of total FPs---this is because AUT($Pr,24m$) is sensitive to the time at which FPs occur; the same observation is valid for total FNs and AUT Recall. After tuning, the $F_1$ performance of \drebin{} and \dl{} become similar, although \dl{} remains  higher in terms of AUT. The tuning improves the AUT($F_1,24m$) of \dl{}  only marginally, as \dl{} is already robust to time decay even before tuning (\autoref{fig:decay}).

The next section focuses on the two classifiers less robust to time decay, \drebin{} and \mmd{}, to evaluate with \tool{} the performance-cost trade-offs of budget-constrained strategies for delaying time decay.

\begin{table}[t]
\centering
\small
  
  \resizebox{0.9\columnwidth}{!}{%
  
  \begin{tabular}{|l|l|r|r||c|c|c|}
\hline
  \multirow{2}{*}{{\bf Algorithm}}&
  \multirow{2}{*}{\textbf{$\varphi$}} &
  \multirow{2}{*}{\textbf{FP}} &
  \multirow{2}{*}{\textbf{FN}} &
  \multicolumn{3}{c|}{\textbf{AUT($\mathbb{P}$,24m)}} \\
  \hhline{~~~~---}

  &
  &
  &
  & $F_1$ & $Pr$ & $Rec$ \\
\hline \hline
  \multirow{4}{*}{\drebin{}~\cite{Arp:Drebin}}
    & 10\% ($\hat{\sigma}$)          & 965  & 3,851 & 0.58 & 0.75 & 0.48\\
    & 25\% ($\varphi^{*}_{F_1}$)           & 2,156 & 2,815 & \cellcolor{lightgray!90} 0.62 & 0.65 & 0.61 \\
    & 10\% ($\varphi^{*}_{Pr}$)           & 965 & 3,851 & 0.58 & \cellcolor{lightgray!90} 0.75 & 0.48 \\
    & 50\% ($\varphi^{*}_{Rec}$)           & 3,728 & 1,793 & 0.64 & 0.58 & \cellcolor{lightgray!90} 0.74  \\
    \hline \hline

  \multirow{4}{*}{\mmd{}~\cite{Mariconti:MaMaDroid}}
    & 10\% ($\hat{\sigma}$)   & 274  & 5,689 &  0.32 & 0.77 & 0.20 \\
    & 50\% ($\varphi^{*}_{F_1}$)    & 4,160  & 2,689 & \cellcolor{lightgray!90} 0.53 & 0.50 & 0.60\\
    & 10\% ($\varphi^{*}_{Pr}$)           & 274 & 5,689 & 0.32 & \cellcolor{lightgray!90} 0.77 & 0.20 \\
    & 50\% ($\varphi^{*}_{Rec}$)    & 4,160  & 2,689 & 0.53 & 0.50 & \cellcolor{lightgray!90} 0.60\\
    
    \hline
    \hline

  \multirow{4}{*}{\dl{}~\cite{Papernot:ESORICS}}
    & 10\% ($\hat{\sigma}$)   & 968  & 3,291 &  0.64 & 0.78 & 0.53 \\
    & 25\% ($\varphi^{*}_{F_1}$)    & 2,284  & 2,346 & \cellcolor{lightgray!90} 0.65 & 0.66 & 0.65\\
    & 10\% ($\varphi^{*}_{Pr}$)           & 968 & 3,291 & 0.64 & \cellcolor{lightgray!90} 0.78 & 0.53 \\
    & 25\% ($\varphi^{*}_{Rec}$)    & 2,284  & 2,346 & 0.65 & 0.66 & \cellcolor{lightgray!90} 0.65\\
    
    \hline

  \end{tabular}
  }
  \caption{Testing AUTs performance over 24 months when training with $\hat{\sigma}$, $\varphi^*_{F_1}$, $\varphi^*_{Pr}$ and $\varphi^*_{Rec}$.}
\label{tab:varphi}
\end{table}

\section{Delaying Time Decay}
\label{sec:delay}

We have shown how enforcing constraints and computing AUT with \tool{} can reveal the real performance of Android malware classifiers (\autoref{sec:theexample}). This \emph{baseline AUT performance} (without retraining) allows users to evaluate the general robustness of an algorithm to time decay.
A classifier may be retrained to update its model. However, \emph{manual labeling} is costly (especially in the Android malware setting), and the ML community~\cite{Settles:AL,Bart:Reject} has worked extensively on mitigation strategies---e.g., to identify a limited number of \emph{best} objects to label (active learning). While effective at postponing time decay, strategies like these can further complicate the fair evaluation and comparison of classifiers.

In this section, we show how \tool{} can be used to compare and
evaluate the trade-offs of different budget-constrained strategies to
delay time decay. Since \dl{} has shown to be more robust to
time decay (\autoref{sec:theexample}) than \drebin{} and \mmd{}, in this
section we focus our attention these to show performance-cost trade-offs of different mitigations.

\begin{figure}[t]
 	\centering
	\includegraphics[width=1\columnwidth]{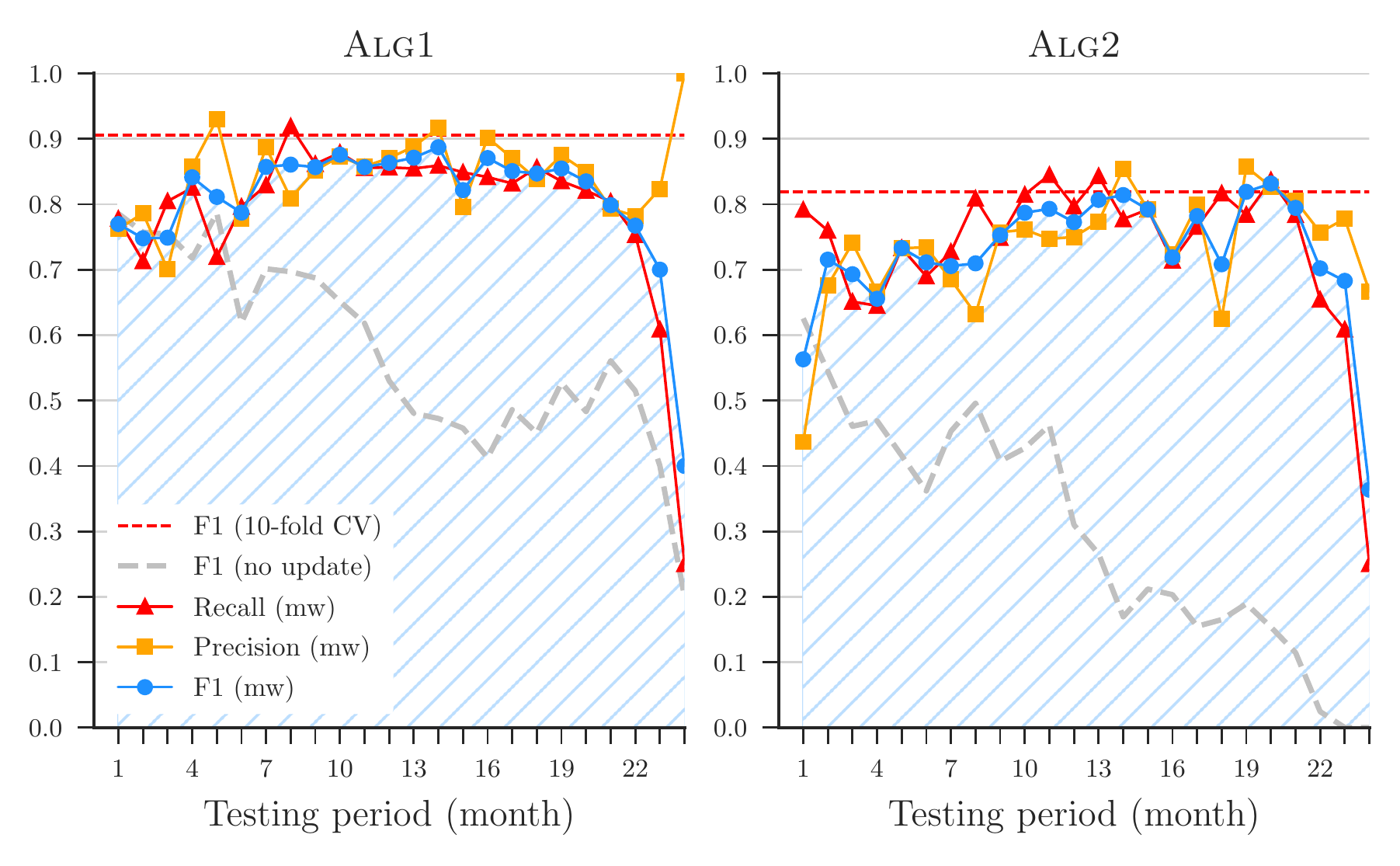}
	\vspace{-15px}
	\caption{Delaying time decay: incremental retraining.}
	\label{fig:incremental}
\end{figure}

\begin{figure*}[t]
	\hspace*{-0.7cm}
	\subfigure[$F_1$-Score]{
	\includegraphics[width=0.33\textwidth]{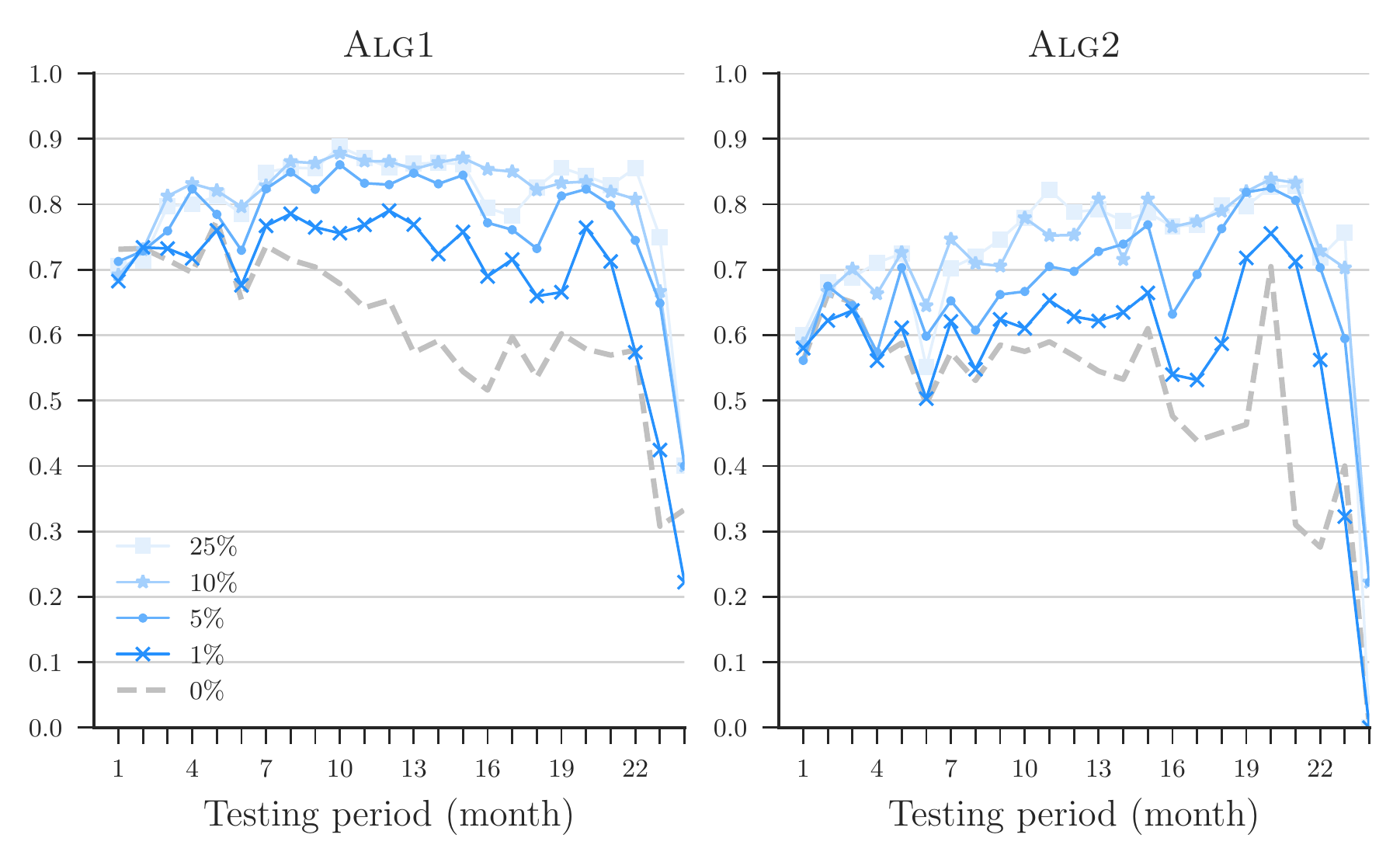}
	}
	\subfigure[Precision]{
	\includegraphics[width=0.33\textwidth]{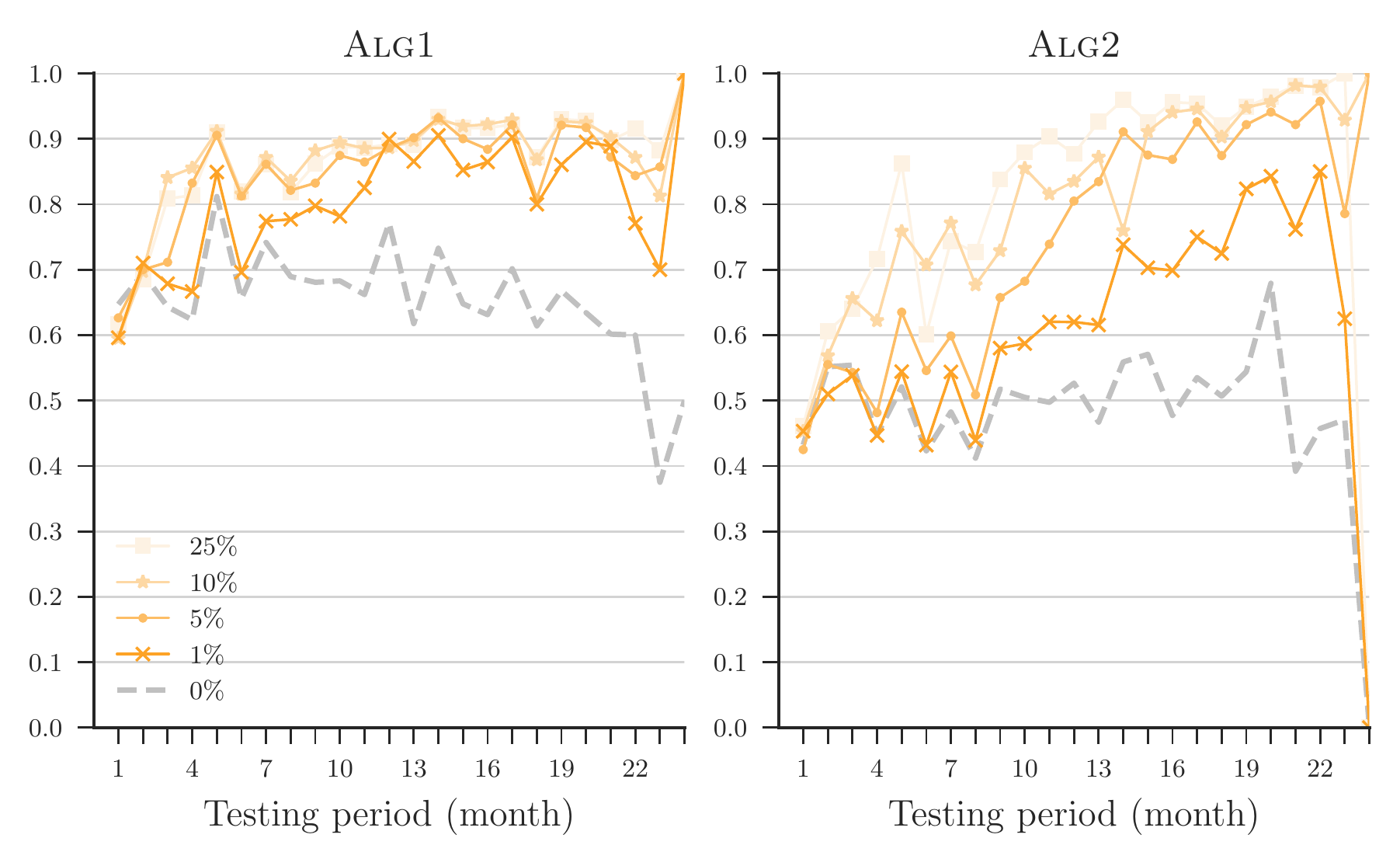}
	}
	\subfigure[Recall]{
	\includegraphics[width=0.33\textwidth]{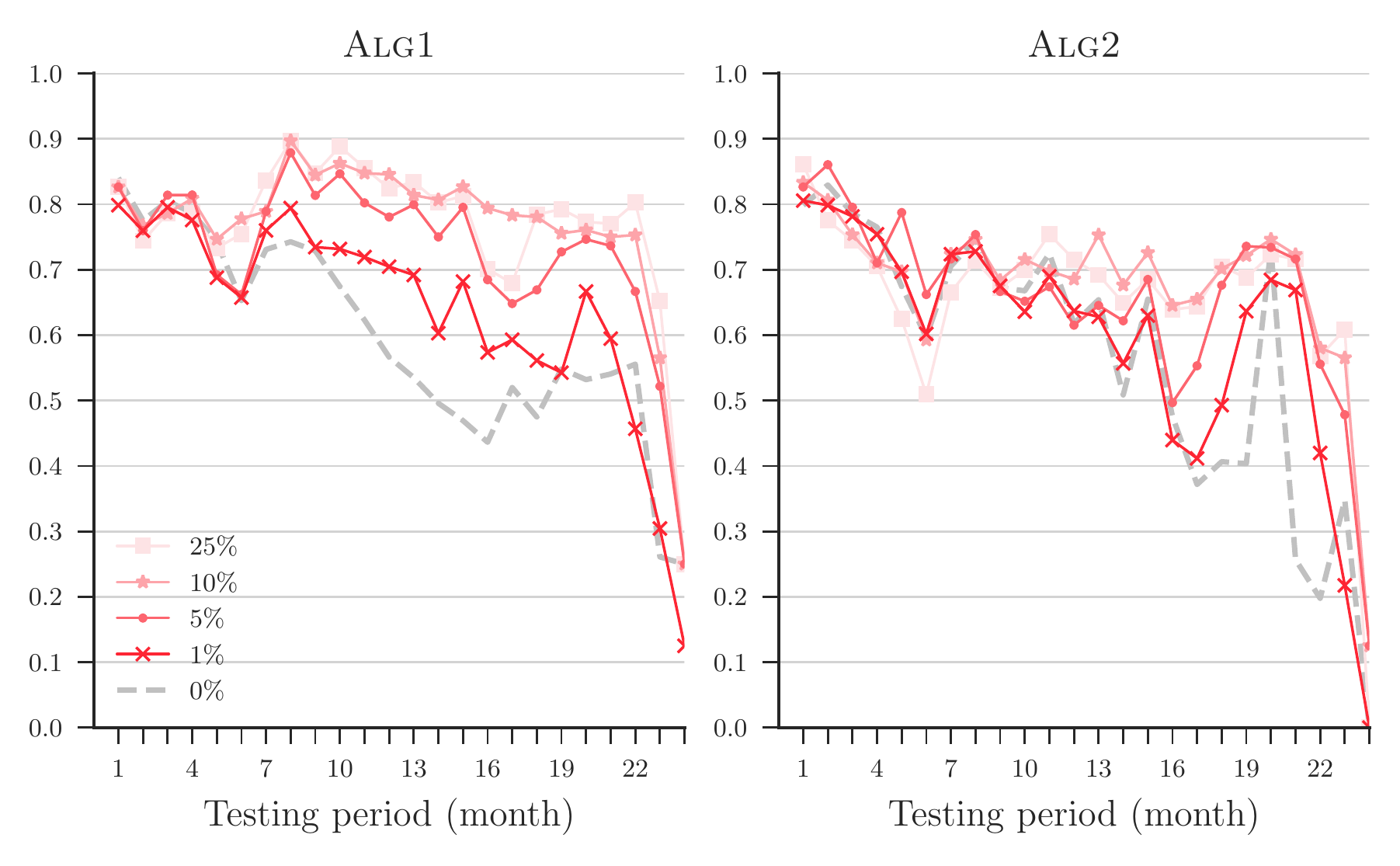}
	}
	\hspace*{-0.7cm}
	\vspace{-10px}
	\caption{Delay time decay: performance with active learning based on uncertainty sampling.}
	\label{fig:al}
\end{figure*}

\begin{figure}[t]
\centering
\includegraphics[width=1\columnwidth]{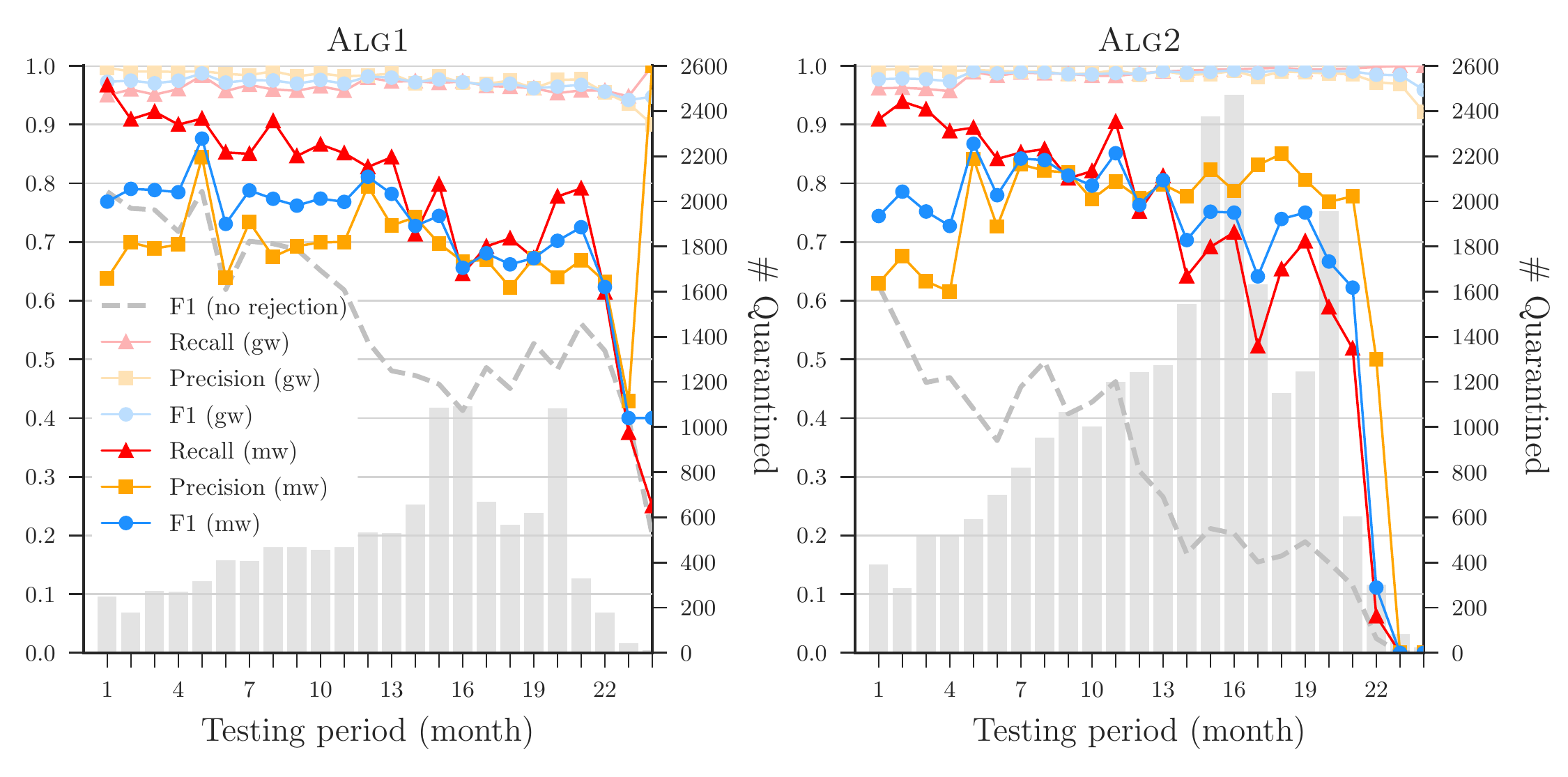}
\vspace{-20px}
\caption{Delay time decay: classification with rejection.}
\label{fig:rejection}
\end{figure}

\subsection{Delay Strategies}

We do not propose novel delay strategies, but instead focus on how \tool{} allows for the comparison of some popular approaches to mitigating time decay. This shows researchers how to adopt \tool{} for the fair comparison of different approaches when proposing novel solutions to delaying time decay under budget constraints. We now summarize the delay strategies we consider and show results on our dataset. For interested readers, we include additional background knowledge on these strategies in~\autoref{sec:delaydetails}.

{\bf Incremental retraining.} We first consider an approach that
represents an ideal upper bound on performance, where \emph{all}
points are included in retraining every month. This is likely
unrealistic as it requires continuously labeling all the objects. Even
assuming a reliance on VirusTotal, there is still an API usage cost
associated with higher query rates and the approach may be ill-suited
in other security domains.  \autoref{fig:incremental} shows the performance of \drebin{} and \mmd{} with monthly incremental retraining.

{\bf Active learning.} Active Learning (AL) strategies investigate how
to select a subset of test objects (with unknown labels) that, if
manually labeled and included in the training set, should be the most
valuable for updating the classification model~\cite{Settles:AL}.
Here, we consider the most popular AL query strategy, \emph{uncertainty sampling}, in which the points with the most uncertain predictions are selected for retraining, under the intuition that they are the most relevant to adjust decision boundaries.
\autoref{fig:al} reports the active learning results obtained with uncertainty sampling, for different percentages of objects labeled per month. We observe that even with 1\% AL, the performance already improves significantly.

{\bf Classification with rejection.} Another mitigation strategy involves rejecting a classifier's decision as ``low confidence'' and delaying the decision to a future date~\cite{Bart:Reject}. This isolates the rejected objects to a \emph{quarantine} area which will later require manual inspection.
\autoref{fig:rejection} reports the performance of \drebin{} and \mmd{} after applying a reject option based on~\cite{Jordaney:Transcend}. In particular, we use the third quartile of probabilities of incorrect predictions as the rejection threshold~\cite{Jordaney:Transcend}. The gray histograms in the background report the number of rejected objects per month. The second year of testing has more rejected objects for both \drebin{} and \mmd{}, although \mmd{} overall rejects more objects.

\begin{table}[t]
\centering
\small
        \resizebox{\columnwidth}{!}{%
  \begin{tabular}{|l||c|c||c|c||c|c||c|c|}
\hhline{~--------}
   \multicolumn{1}{c||}{} &
  \multicolumn{4}{c||}{\textbf{Costs}} &
    \multicolumn{4}{c|}{\textbf{Performance}} \\ \hhline{~--------}
 \multicolumn{1}{c||}{} &
  \multicolumn{2}{c||}{\multirow{2}{*}{\labelingcost{}}} &
    \multicolumn{2}{c||}{\multirow{2}{*}{\quarantinecost{}}} &
    \multicolumn{4}{c|}{\performance{} : AUT($F_1,24m$)} \\
\hhline{-~~~~----}

  \multirow{1}{*}{\bf Delay} &
  \multicolumn{2}{c||}{} &
    \multicolumn{2}{c||}{} &
    \multicolumn{2}{c|}{$\varphi=\hat{\sigma}$} &
    \multicolumn{2}{c|}{$\varphi=\varphi^*_{F_1}$} \\
\hhline{~--------}

   {\bf method} &
  \drebin{} & \mmd{} &
  \drebin{} & \mmd{} &
  \drebin{} & \mmd{} & \drebin{} & \mmd{} \\ \hline \hline

No update &0&0&0&0&0.577& 0.317& \cellcolor{orange!10}0.622&\cellcolor{blue!10}  0.527\\ \hline \hline
Rejection ($\hat{\sigma}$) &0&  0&10,283&3,595&\cellcolor{orange!10}0.717&0.280 & -- & -- \\ \hline
Rejection ($\varphi^*_{F_1}$) &0&  0&10,576&24,390&--&--&0.704& \cellcolor{blue!10} 0.683\\ \hline
AL: 1\%&709&709&0&0& \cellcolor{orange!10} 0.708 & 0.456 &0.703& \cellcolor{blue!10}  0.589\\ \hline
AL: 2.5\%&1,788&1,788&0&0& 0.738 & 0.509 & \cellcolor{orange!10} 0.758& \cellcolor{blue!10} 0.667\\ \hline
AL: 5\%&3,589&3,589&0&0& 0.782 & 0.615 & \cellcolor{orange!10}0.784& \cellcolor{blue!10} 0.680\\ \hline
AL: 7.5\%&5,387&5,387&0&0& 0.793 & 0.641 & \cellcolor{orange!10}0.801& \cellcolor{blue!10} 0.714\\ \hline
AL: 10\%&7,189&7,189&0&0& 0.796 & 0.656 & \cellcolor{orange!10}0.802&\cellcolor{blue!10} 0.732\\ \hline
AL: 25\%&17,989&17,989&0&0& 0.821 & 0.674 & \cellcolor{orange!10}0.823&\cellcolor{blue!10} 0.732\\ \hline
AL: 50\%&35,988&35,988&0&0& 0.817 & 0.679 &\cellcolor{orange!10} 0.828&\cellcolor{blue!10} 0.741\\ \hline
Inc. retrain&71,988&71,988&0&0& 0.818 & 0.679 &\cellcolor{orange!10} 0.830&\cellcolor{blue!10} 0.736\\ \hline
  \end{tabular}%
  }

  \caption{Performance-cost comparison of delay methods.}
  \label{tab:pc-tradeoff}

\end{table}

\subsection{Analysis of Delay Methods}

To quantify performance-cost trade-offs of methods to delay time decay without changing the algorithm, we characterize the following three elements: {\bf Performance} (\performance{}), the performance measured in terms of AUT to capture robustness against time decay~(\autoref{sec:aut}); {\bf Labeling Cost} (\labelingcost{}), the number of testing objects (if any) that must be labeled---the labeling must occur periodically (e.g., every month), and is particularly costly in the malware domain as manual inspection requires many resources (infrastructure, time, expertise, etc)---for example, Miller et al.~\cite{Miller:Reviewer} estimated that an average company could manually label 80 objects per day; {\bf Quarantine Cost} (\quarantinecost{}), the number of objects (if any) rejected by the classifier---these must be manually verified, so there is a cost for leaving them in quarantine.

\autoref{tab:pc-tradeoff}, utilizing AUT($F_1$,24m) while enforcing our constraints, reports a summary of labeling cost \labelingcost{}, quarantine cost \quarantinecost{}, and two performance columns \performance{}, corresponding to training with $\hat{\sigma}$ and $\varphi^*_{F_1}$ (\autoref{sec:tuning}), respectively. In each row, we highlight in purple cells (resp. orange) the column with the highest AUT for \mmd{} (resp. \drebin{}). \autoref{tab:pc-tradeoff} allows us to: (i) examine the effectiveness of the training ratios~$\varphi^*_{F_1}$ and~$\hat{\sigma}$; (ii) analyze the AUT performance improvement and the corresponding costs for delaying time decay; (iii) compare the performance of \drebin{} and \mmd{} in different settings.

First, let us compare $\varphi^*_{F_1}$ with $\hat{\sigma}$. The first row of
\autoref{tab:pc-tradeoff} represents the scenario in which the model is trained
only once at the beginning---the scenario for which we originally designed
Algorithm~\ref{alg:varphi} (\autoref{sec:tuning} and \autoref{fig:varphi}). Without
methods to delay time decay, $\varphi^*_{F_1}$ achieves better performance than
$\hat{\sigma}$ for both \drebin{} and \mmd{} at no cost. In all other
configurations, we observe that training $\varphi=\varphi^*_{F_1}$ always
improves performance for \mmd{}, whereas for \drebin{} it is slightly
advantageous in most cases except for rejection and AL 1\%---in general,
the performance of \drebin{} trained with $\varphi^*_{F_1}$ and $\hat{\sigma}$ is
consistently close.
The intuition for this outcome is that $\varphi^*_{F_1}$ and $\hat{\sigma}$ are
also close for \drebin{}: when applying the AL strategy, we re-apply Algorithm~\ref{alg:varphi} at each
step and find that the average
$\varphi^*_{F_1} \approx 15\%$ for \drebin{}, which is close to 10\% (i.e.,
$\hat{\sigma}$). On the other hand, for \mmd{} the average
$\varphi^*_{F_1} \approx 50\%$, which is far from $\hat{\sigma}$ and improves
all results significantly. We can conclude that our tuning algorithm is most effective when it finds a $\varphi^*_\mathbb{P}$ that differs
from the estimated $\hat{\sigma}$.

Then, we analyze the performance improvement and related cost of using delay
methods. The improvement in $F_1$-Score granted by our algorithm comes at no
labeling or quarantine cost. We can observe that one can improve the
in-the-wild performance of the algorithms at some cost \labelingcost{} or
\quarantinecost{}. It is important to observe that objects discarded or to be
labeled are not necessarily malware; they are just the objects most uncertain
according to the algorithm, which the classifier may have likely
misclassified. The labeling costs \labelingcost{} for \drebin{} and \mmd{} are
identical (same dataset); in
AL, the percentage of retrained objects is user-specified and fixed.

Finally, \autoref{tab:pc-tradeoff} shows that \drebin{} consistently outperforms \mmd{} on $F_1$ for all performance-cost trade-offs. This confirms the trend seen in the realistic settings of~\autoref{tab:bias}.

This section shows that \tool{} is helpful to both researchers and industrial practitioners. Practitioners need to estimate the performance of a classifier in the wild, compare different algorithms, and determine resources required for \labelingcost{} and \quarantinecost{}. For researchers, it is useful to understand how to reduce costs \labelingcost{} and \quarantinecost{} while improving  classifiers performance \performance{} through comparable, unbiased evaluations. The problem is challenging, but we hope that releasing \tool{}'s code fosters further research and widespread adoption.

\section{Discussion}
\label{sec:discussion}

We now discuss guidelines, our assumptions, and how we address limitations of our work.

\textbf{Actionable points on \tool{}.} It is relevant to discuss how both
researchers and practitioners can benefit from \tool{} and our
findings. A \emph{baseline AUT performance} (without classifier
retraining) allows users to evaluate the general robustness of an
algorithm to performance decay (\autoref{sec:aut}). We demonstrate how
\tool{} can reveal true performance and provide counter-intuitive
results (\autoref{sec:theexample}). Robustness over extended time periods
is practically relevant for deployment scenarios without the financial
or computational resources to label and retrain often. Even with
retraining strategies (\autoref{sec:delay}), classifiers may not perform
consistently over time. Manual labeling is costly, and the ML
community has worked on mitigation strategies to identify a limited
number of \emph{best} objects to label (e.g., active
learning~\cite{Settles:AL}). \tool{} takes care of removing
spatio-temporal bias from evaluations, so that researchers can focus
on the proposal of more robust algorithms (\autoref{sec:delay}). In this
context, \tool{} allows for the creation of comparable baselines for
algorithms in a time-aware setting. Moreover, \tool{} can be used with
different time granularity, provided each period has a significant
number of samples. For example, if researchers are interested in
increasing robustness to decay for the upcoming 3 months, they can use \tool{} to produce bias-free comparisons of their approach with prior research, while considering time decay.

{\bf Generalization to other security domains.}
Although we used \tool{} in the Android domain, our methodology generalizes and can be immediately applied to any machine learning-driven security domain to achieve an evaluation without spatio-temporal bias. Our methodology, although general, requires some domain-specific parameters that reflect realistic conditions (e.g., time granularity $\Delta$ and test time length). This is not a weakness of our work, but rather an expected requirement.
In general, it is reasonable to expect that spatio-temporal bias may afflict other security domains when affected by concept drift and i.i.d. does not hold---however, further experiments in other domains (e.g., Windows malware, code vulnerabilities) are required to make any scientific conclusion. \tool{} can be used to understand the extent to which spatio-temporal bias affects such security domains; however, the ability to generalize requires access to large timestamped datasets, knowledge of realistic class ratios, and code or sufficient details to reproduce baselines.

{\bf Domain-specific in-the-wild malware percentage~$\hat{\sigma}$.}
In the Android landscape, we assume that $\hat{\sigma}$ is around 10\% (\autoref{sec:tsdist}). Correctly estimating the malware percentage in the testing dataset is a challenging task and we encourage further representative measurement studies~\cite{lindorfer2014andradar,Perdisci:Measuring} and data sharing to obtain realistic experimental settings.

{\bf Correct observation labels.} We assume goodware and malware labels in the dataset are correct (\autoref{sec:dataset}).
Miller et al.~\cite{Miller:Reviewer} found that AVs sometimes change their outcome over time: some goodware may eventually be tagged as malware.  However, they also found that VirusTotal detections stabilize after one year; since we are using observations up to Dec 2016, we consider VirusTotal's labels as reliable. In the future, we may integrate approaches for \emph{noisy oracles}~\cite{Du:NoisyAL}, which assume some observations are mislabeled.

{\bf Timestamps in the dataset.} It is important to consider that some timestamps in a public dataset could be incorrect or invalid. In this paper, we rely on the public AndroZoo dataset maintained at the University of Luxembourg, and we rely on the {\tt dex\_date} attribute as the approximation of an observation timestamp, as recommended by the dataset creators~\cite{Allix:AndroZoo}. We further verified the reliability of the {\tt dex\_date} attribute by re-downloading VirusTotal~\cite{VT:URL} reports for 25K apps\footnote{We downloaded only 25K VT reports (corresponding to about 20\% of our dataset) due to restrictions on our VirusTotal API usage quota.} and verifying that the {\tt first\_seen} attribute always matched the {\tt dex\_date} within our time span.
In general, we recommend performing some sanitization of a timestamped dataset before performing any analysis on it: if multiple timestamps are available for each object, consider the most reliable timestamp you have access to (e.g., the timestamp recommended by the dataset creators, or the VirusTotal's {\tt first\_seen} attribute) and discard objects with ``impossible'' timestamps (e.g., with dates which are either too old or in the future), which may be caused by incorrect parsing or invalid values of some timestamps. To improve trustworthiness of the timestamps, one could verify whether a given object contains time inconsistencies or features not yet available when the app was released~\cite{Li:MoonlightBox}. We encourage the community to promptly notify dataset maintainers of any date inconsistencies.
 In the \tool{}'s project website (\autoref{sec:availability}), we will maintain an updated list of timestamped datasets publicly available for the security community.

{\bf Choosing time granularity ($\Delta$).} Choosing the length of the time slots (i.e., time granularity) largely depends on the sparseness of the available dataset: in general, the granularity should be chosen to be as small as possible, while containing a statistically significant number of samples---as a rule of thumb, we keep the buckets large enough to have at least 1000 objects, which in our case leads to a monthly granularity. If there are restrictions on the number of time slots that can be considered (perhaps due to limited processing power), a coarser granularity can be used; however if the granularity becomes too large then the true trend might not be captured.

{\bf Resilience of malware classifiers.} In our study, we analyze three recent high-profile classifiers. One could argue that other classifiers may show consistently high performance even with space-time bias eliminated. And this should indeed be the goal of research on malware classification. \tool{} provides a mechanism for an unbiased evaluation that we hope will support this kind of work.

{\bf Adversarial ML.} Adversarial ML focuses on perturbing training or testing observations to compel a classifier to make incorrect predictions~\cite{Biggio:Wild}. Both relate to concepts of \textit{robustness} and one can characterize adversarial ML as an artificially induced worst-case concept drift scenario. While the adversarial setting remains an open problem, the experimental bias we describe in this work---endemic in Android malware classification---must be addressed prior to realistic evaluations of adversarial mitigations.

\section{Related Work}
\label{sec:related}

A common experimental bias in security is the \emph{base rate fallacy}~\cite{Axelsson:BaseRate}, which states that in highly-imbalanced datasets (e.g., network intrusion detection, where most traffic is benign), TPR and FPR are misleading performance metrics, because even $\mathrm{FPR}=1\%$ may correspond to \emph{millions} of FPs and only \emph{thousands} of TPs. In contrast, our work identifies experimental settings that are misleading \emph{regardless} of the adopted metrics, and that remain incorrect even if the right metrics are used (\autoref{sec:theexample}).
Sommer and Paxson~\cite{Sommer:Outside} discuss challenges and guidelines in ML-based intrusion detection; Rossow et al.~\cite{rossow2012prudent} discuss best practices for conducting malware experiments; van der Kouwe et al.~\cite{Van:BenchmarkingCrimes} identify 22 common errors in system security evaluations. While helpful, these works~\cite{Van:BenchmarkingCrimes,rossow2012prudent, Sommer:Outside} do not identify temporal and spatial bias, do not focus on Android, and do not \emph{quantify} the impact of errors on classifiers performance, and their guidelines would not prevent all sources of temporal and spatial bias we identify.
To be precise, Rossow et al.~\cite{rossow2012prudent} evaluate the percentage of objects---in previously adopted datasets---that are ``incorrect'' (e.g., goodware labeled as malware, malfunctioning malware), but without evaluating impact on classifier performance.
Zhou et al.~\cite{Zhou:HPC} have recently shown that Hardware Performance Counters (HPCs) are not really effective for malware classification; while interesting and in line with the spirit of our work, their focus is very narrow, and they rely on 10-fold CV in the evaluation.

Allix et al.~\cite{Allix:Timeline} broke new ground by evaluating malware classifiers in relation to time and showing how future knowledge can inflate performance, but do not propose any solution for comparable evaluations and only identify C1. As a separate issue, Allix et al.~\cite{Allix2016} investigated the difference between in-the-lab and in-the-wild scenarios and found that the greater presence  of goodware leads to lower performance. We systematically analyze and explain these issues and help address them by formalizing a set of constraints (jointly considering the impact of temporal and spatial bias), introducing AUT as a unified performance metric for fair time-aware comparisons of different solutions, and offering a tuning algorithm to leverage the effects of training data distribution.
Miller et al.~\cite{Miller:Reviewer} identified \emph{temporal sample consistency} (equivalent to our constraint C1), but not C2 or C3---which are fundamental (\autoref{sec:theexample}); moreover, they considered the test period to be a uniform time slot, whereas we take time decay into account.
Roy et al.~\cite{Roy:ExpML} questioned the use of recent or older malware as training objects and the performance degradation in testing real-world object ratios; however, most experiments were designed without considering time, reducing the reliability of their conclusions.
While past work highlighted some sources of experimental bias~\cite{Miller:Reviewer,Roy:ExpML,Allix:Timeline,Allix2016}, it also gave little consideration to classifiers' aims: different scenarios may have different goals (not necessarily maximizing $F_1$), hence in our work we show the effects of different training settings on performance goals and propose an algorithm to properly tune a classifier accordingly (\autoref{sec:tuning}).

Other works from the ML literature investigate imbalanced datasets and highlighted how training and testing ratios can influence the results of an algorithm~\cite{Weiss:ReBalance,He:Imbalanced,chawla2004special}. However, not coming from the security domain, these studies~\cite{Weiss:ReBalance,He:Imbalanced,chawla2004special} focus only on some aspects of spatial bias and do \emph{not} consider temporal bias. Indeed, concept drift is less problematic in some applications (e.g., image and text classification) than in Android malware~\cite{Jordaney:Transcend}. Fawcett~\cite{Fawcett:Spam} focuses on challenges in spam detection, one of which resembles spatial bias; no solution is provided, whereas we introduce C3 to this end and demonstrate how its violation inflates performance (\autoref{sec:theexample}). Torralba and Efros~\cite{Torralba:Unbiased} discuss the problem of \emph{dataset bias} in computer vision, distinct from our security setting where there are fewer benchmarks; moreover in images the negative class (e.g., ``not cat'') can grow arbitrarily, which is less likely in the malware context. Moreno-Torres et al.~\cite{Moreno:Unifying} systematize different \emph{drifts}, and mention \emph{sample-selection bias}; while this resembles spatial bias, they do not propose any solution/experiments for its impact on ML performance. Other related work underlines the importance of choosing appropriate performance metrics to avoid an incorrect interpretation of the results (e.g., ROC curves are misleading in an imbalanced dataset~\cite{hand2009measuring,davis2006relationship}). In this paper, we take imbalance into account, and we propose actionable constraints and metrics with tool support to evaluate performance decay of classifiers over time.

{\bf Summary.} Several studies of bias exist and have motivated our research. However, none of them address the entire problem in the context of evolving data (where the i.i.d. assumption does not hold anymore). Constraint C1, introduced by Miller et al.~\cite{Miller:Reviewer}, is by itself insufficient to eliminate bias. This is evident from the original evaluation in {\sc MaMaDroid}~\cite{Mariconti:MaMaDroid}, which enforces only C1. The evaluation in~\autoref{sec:theexample} clarifies why our novel constraints C2 and C3 are fundamental, and shows how our AUT metric can effectively reveal the true performance of algorithms, providing counter-intuitive results.

\section{Availability}
\label{sec:availability}

We make \tool{}'s code and data available to the research community to promote the adoption of a sound and unbiased evaluation of classifiers. The \tool{} project website with instructions to request access is at \mbox{\sf https://s2lab.kcl.ac.uk/projects/tesseract/}. We will also maintain an updated list of publicly available security-related datasets with timestamped objects.

\section{Conclusions}
\label{sec:conclusions}

We have identified novel temporal and spatial bias in the Android domain and proposed novel constraints, metrics and tuning to address such issues. We have built and released  \tool{} as an open-source tool that integrates our methods. We have shown how \tool{} can reveal the real performance of malware classifiers that remain hidden in wrong experimental settings in a non-stationary context. \tool{} is fundamental for the correct evaluation and comparison of different solutions, in particular when considering mitigation strategies for time decay. We are currently working on integrating a time-varying percentage of malware in our framework to model still more realistic scenarios, and on how to use the slope of the performance decay curve to better differentiate algorithms with similar AUT. 

We envision that future work on Android malware classification will use \tool{} to produce realistic, comparable and unbiased results. Moreover, we also encourage the security community to adopt \tool{} to evaluate the impact of temporal and spatial bias in other security domains where concept drift still needs to be quantified.

\section*{Acknowledgements}

We thank the anonymous reviewers and our shepherd, Roya Ensafi, for their constructive feedback, which has improved the overall quality of this work. This research has been partially sponsored by the UK EP/L022710/1 and EP/P009301/1 EPSRC research grants.

%{\small
%\bibliographystyle{plain}
%\input{main.bbl}
%\bibliography{references}}

\begin{appendix}
\section{Appendix}

\subsection{Algorithm Hyperparameters}
\label{sec:hyperparam}

We hereby report the details of the hyperparameters used to replicate \drebin{}, \mmd{} and \dl{}.

We replicate the settings and experiments of \drebin{}~\cite{Arp:Drebin} (linear SVM with C=1) and \mmd{}~\cite{Mariconti:MaMaDroid} (package mode and RF with $101$ trees and max depth of $64$) as described in the respective papers~\cite{Arp:Drebin,Mariconti:MaMaDroid}, successfully reproducing the published results. Since on our dataset the \drebin{} performance is slightly lower (around 0.91 10-fold $F_1$), we also reproduce the experiment on their same dataset~\cite{Arp:Drebin}, achieving their original performance of about 0.94 10-fold $F_1$. We have used \sklearn{}, with {\it sklearn.svm.LinearSVC} for~\drebin{} and {\it sklearn.ensemble.RandomForestClassifier} for~\mmd{}.

We then follow the guidelines in~\cite{Papernot:ESORICS} to re-implement \dl{} with \keras{}. The features given as initial input to the neural network are the same as \drebin{}. We replicated the best-performing neural network architecture of~\cite{Papernot:ESORICS}, by training with 10 epochs and batch size equal to 1,000. To perform the training optimization, we used the stochastic gradient descent class \textit{keras.optimizers.SGD} with the following parameters: \emph{lr}=0.1, \emph{momentum}=0.0, \emph{decay}=0.0, \emph{nesterov}=False.
Some low-level details of the hyperparameter optimization were missing
from the original paper~\cite{Papernot:ESORICS}; we managed to obtain
slightly higher $F_1$ performance in 10-fold setting
(\autoref{sec:theexample}) likely because they have performed hyperparameter optimization on the Accuracy metric~\cite{Bishop:ML}---which is misleading in imbalanced datasets~\cite{Axelsson:BaseRate} where one class is prevalent (goodware, in Android).

\subsection{Symbol table}
\label{app:symbol}

\autoref{tab:symbol} is a legend of the main symbols used throughout this paper
to improve readability.

\begin{table}[h!]
\centering
\small
{
\footnotesize
  \begin{tabular}{m{10mm}|m{70mm}}
\centering
  \textbf{Symbol} & \textbf{Description} \\
  \midrule
  gw & Short version of goodware. \\  \hline
  mw & Short version of malware. \\  \hline
  ML & Short version of Machine Learning. \\  \hline
  \dataset{} & Labeled dataset with malware (mw) and goodware (gw). \\  \hline
  $\mathit{Tr}$ & Training dataset. \\  \hline
  $W$ & Size of the time window of the training set (e.g., 1 year). \\  \hline
  $\mathit{Ts}$ & Testing dataset. \\  \hline
  $S$ & Size of the time window of the testing set (e.g., 2 years). \\  \hline
  $\Delta$ & Size of the test time-slots for time-aware evaluations (e.g., months). \\  \hline
  AUT($f$,$N$) & Area Under Time, a new metric we define to measure performance over time decay and compare different solutions (\autoref{sec:aut}). It is always computed with respect to a performance function $f$ (e.g., $F_1$-Score) and $N$ is the number of time units considered (e.g., 24 months)\\ \hline
  $\hat{\sigma}$ & Estimated percentage of malware (mw) in the wild. \\ \hline
  $\varphi$ & Percentage of malware (mw) in the training set. \\  \hline
  $\delta$ & Percentage of malware (mw) in the testing set. \\  \hline
  $\mathbb{P}$ & Performance target of the tuning algorithm in~\autoref{sec:tuning}; it can be $F_1$-Score, Precision ($Pr$) or Recall ($Rec$). \\ \hline
  $\varphi^*_\mathbb{P}$ & Percentage of malware (mw) in the training set, to improve performance $\mathbb{P}$ on the malware (mw) class (\autoref{sec:tuning}). \\  \hline
\errorrate{} & Error rate (\autoref{sec:tuning}). \\  \hline
\errorratemax{} & Maximum error rate when searching $\varphi^*_\mathbb{P}$ (\autoref{sec:tuning}). \\  \hline
 $\Theta$ & Model learned after training a classifier. \\  \hline
  \labelingcost{} & Labeling cost. \\  \hline
 \quarantinecost{} & Quarantine cost. \\  \hline
 \performance{} & Performance; depending on the context, it will refer to AUT with $F_1$ or $Pr$ or $Rec$. \\  \hline
  \bottomrule
  \end{tabular}
}
\caption{Symbol table.}
\label{tab:symbol}

\end{table}

\subsection{Cumulative Plots for Time Decay}
\label{sec:app:cml}

\begin{figure*}[t]
	\centering
	\includegraphics[width=0.9\textwidth]{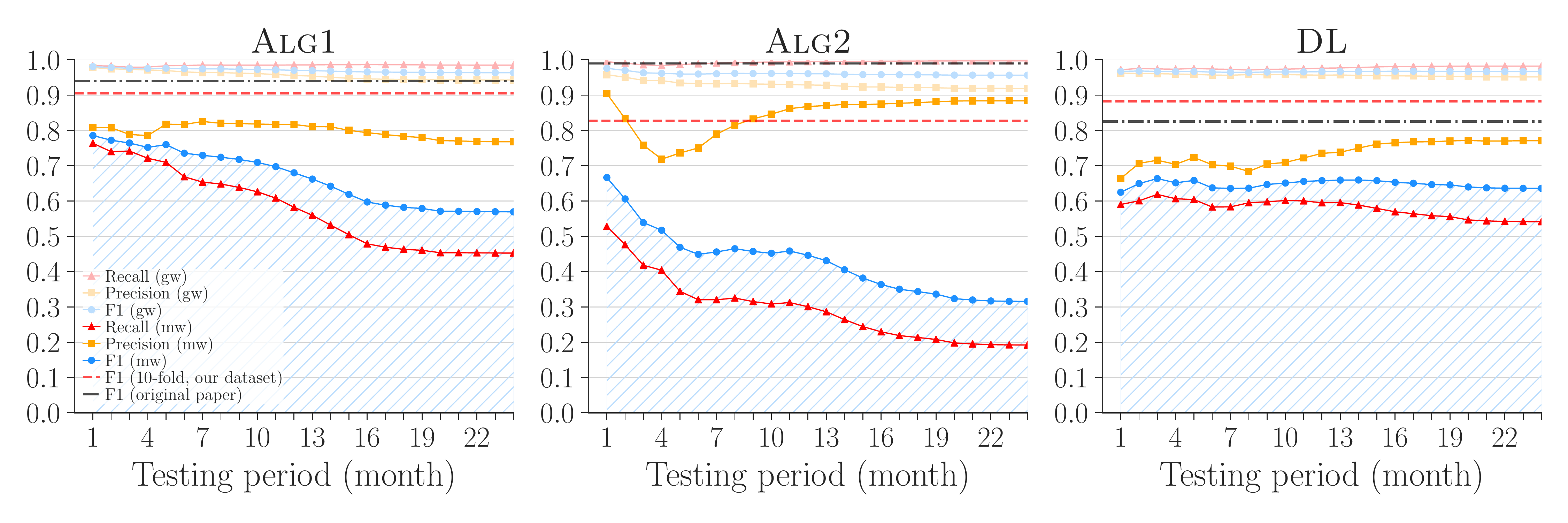}
	\vspace{-10px}
	\caption{Performance time decay with cumulative estimate for \drebin{}, \mmd{} and \dl{}. Testing distribution has $\delta=10\%$ malware, and training distribution has $\varphi=10\%$ malware. }
	\label{fig:decay-cum}
\end{figure*}

\autoref{fig:decay-cum} shows the cumulative performance plot defined in~\autoref{sec:aut}. This is the cumulative version of~\autoref{fig:decay}.

\subsection{Delay Strategies}
\label{sec:delaydetails}

We discuss more background details on the mitigation strategies adopted in Section~\ref{sec:delay}.

{\bf Incremental retraining.}
Incremental retraining is an approach that tends towards an ``ideal'' performance \performancestar{}: \textit{all} test objects are periodically labeled manually, and the new knowledge introduced to the classifier via retraining.  More formally, the performance of month $m_i$ is determined from the predictions of a model~$\Theta$ trained on: $Tr \cup \{m_0,m_1,...,m_{i-1}\}$, where $\{m_0,m_1,...,m_{i-1}\}$ are testing objects, which are manually labeled. The dashed gray line represents the $F_1$-Score \emph{without} incremental retraining (i.e., stationary training). Although incremental retraining generally achieves optimal performance throughout the whole test period, it also incurs the highest labeling cost \labelingcost{}.

{\bf Active learning.}
Active learning is a field of machine learning that studies \emph{query strategies} to select a small number of testing points close to the decision boundaries, that, if included in the training set, are the most relevant for updating the classifier. For example, in a linear SVM the slope of the decision boundary greatly depends on the points that are closest to it, the \emph{support vectors}~\cite{Bishop:ML}; all the points further from the SVM decision boundary are classified with higher confidence, hence have limited effect on the slope of the hyperplane.

We evaluate the impact of one of the most popular active learning strategies: \emph{uncertainty sampling}~\cite{Settles:AL,Miller:Reviewer}. This query strategy selects the most points the classifier is least certain about, and uses them for retraining; we apply it in a time-aware scenario, and choose a percentage of objects to retrain per month. The intuition is that the most uncertain elements are the ones that may be indicative of concept drift, and new, correct knowledge about them may better inform the decision boundaries. The number of objects to label depends on the user's available resources for labeling.

More formally, in binary classification uncertainty sampling gives a score $x^{*}_{LC}$ (where LC stands for \emph{Least Confident}) to each sample~\cite{Settles:AL}; this score is defined as follows\footnote{In multi-class classification, there is a query strategy based on the entropy of the prediction scores array; in binary classification, the entropy-based query strategy is proven to be equivalent to the ``least confident''~\cite{Settles:AL}.}:
\begin{align}
		x^{*}_{LC} := \text{argmax}_x \{ 1 - P_\Theta (\hat{y}|x) \}
		\label{eq:uncertainty}
\end{align}
where $\hat{y} := \text{argmax}_y P_\Theta (y|x)$ is the class label
with the highest posterior probability according to classifier
$\Theta$. In a binary classification task, the maximum uncertainty for
an object is achieved when its prediction probability equal to $0.5$
for both classes (i.e., equal probability of being goodware or
malware). The test objects are sorted by descending order of uncertainty $x^{*}_{LC}$, and the top-n most uncertain are selected to be labeled for retraining the classifier.

Depending on the percentage of manually labeled points, each scenario corresponds to a different labeling cost \labelingcost{}. The labeling cost \labelingcost{} is known a priori since it is user specified.

{\bf Classification with rejection.}
Malware evolves rapidly over time, so if the classifier is not up to
date, the decision region may no longer be representative of new
objects. Another approach, orthogonal to active learning, is to
include a \emph{reject option} as a possible classifier
outcome~\cite{Fumera:Rejection,Jordaney:Transcend}. This discards
the most uncertain predictions to a \emph{quarantine} area for manual
inspection at a future date. At the cost of rejecting some objects,
the overall performance of the classifier (on the remaining objects)
increases. The intuition is that in this way only high confidence
decisions are taken into account.
Again, although performance \performance{} improves, there is a quarantine cost \quarantinecost{} associated with it; in this case, unlike active learning, the cost is not known a priori because, in traditional classification with rejection, a threshold on the classifier confidence is applied~\cite{Fumera:Rejection,Jordaney:Transcend}.

\subsection{\tool{} Implementation}
\label{sec:artifact}

We have implemented our constraints, metrics, and algorithms as a Python library named \tool{}, designed to integrate easily with common workflows. In particular, the API design of \tool{} is heavily inspired by and fully compatible with \sklearn{}~\cite{scikit-learn} and \keras{}~\cite{KERAS}; as a result, many of the conventions and workflows in \tool{} will be familiar to users of these libraries. Here we present an overview of the library's core modules while further details of the design can be found in~\cite{tesseract:poster}.

{\textbf{temporal.py}} While machine learning libraries commonly involve working with a set of predictors \lstinline{X} and a set of output variables \lstinline{y}, \tool{} extends this concept to include an array of \lstinline{datetime} objects \lstinline{t}. This allows for operations such as the time-aware partitioning of datasets (e.g., \lstinline{time_aware_partition()} and \lstinline{time_aware_train_test_split()}) while respecting temporal constraints C1 and C2.

{\textbf{spatial.py}} This module allows the user to alter the proportion of the positive class in a given dataset. \lstinline{downsample_set()} can be used to simulate the natural class distribution $\hat{\sigma}$ expected during deployment or to tune the performance of the model by over-representing a class during training. To this end we provide an implementation of Algorithm~\ref{alg:varphi} for finding the optimal training proportion $\varphi^*$ (\lstinline{search_optimal_train_ratio()}). This module can also assert that constraint C3 (\autoref{sec:constraints}) has not been violated.

{\textbf{metrics.py}} As \tool{} aims to encourage comparable and reproducible evaluations, we include functions for visualizing classifier assessments and deriving metrics such as the accuracy or total errors from slices of a time-aware evaluation. Importantly we also include \lstinline{aut()} for computing the AUT for a given metric ($F_1$, Precision, AUC, etc.) over a given time~period.

{\textbf{evaluation.py}} Here we include the \lstinline{predict()} and \lstinline{fit_predict_update()} functions that accept a classifier, dataset and set of parameters (as defined in~\autoref{sec:constraints}) and return the results of a time-aware evaluation performed across the chosen periods.

{\textbf{selection.py \textnormal{\textit{and}} rejection.py}} For extending the evaluation to testing model update strategies, these modules provide hooks for novel query and reject strategies to be easily plugged into the evaluation cycle. We already implement many of the methods discussed in~\autoref{sec:delay} and include them with our release. We hope this modular approach lowers the bar for follow-up research in these areas.

\subsection{Summary of Datasets Evaluated by Prior Work}
\label{app:datasets}

As a reference, \autoref{tab:datasets} reports the composition of the dataset used in our paper (1st row) and of the datasets used for experimentally biased evaluations in prior work \drebin{}~\cite{Arp:Drebin}, \mmd{}~\cite{Mariconti:MaMaDroid}, and \dl{}~\cite{Papernot:ESORICS} (2nd and 3rd row). In this paper, we always evaluate \drebin{}, \mmd{} and \dl{} with the dataset in the first row (more details in \autoref{sec:dataset} and \autoref{fig:dataset}), because we have built it to allow experiments without spatio-temporal bias by enforcing constraints C1, C2 and C3. Details on experimental settings that caused spatio-temporal bias in prior work are described in~\autoref{sec:bias} and~\autoref{sec:rules}. \emph{We never use the datasets of prior work~\cite{Arp:Drebin,Mariconti:MaMaDroid,Papernot:ESORICS} in our experiments.}

\begin{table}[h!]
\centering
\small
    \resizebox{1\columnwidth}{!}{%
    \renewcommand{\arraystretch}{1.1}
\begin{tabular}{|l|l|c|r|r||c|}
\hline
\textbf{Work}     & \textbf{Apps}          & \textbf{Date Range} & \multicolumn{1}{c|}{\textbf{\# Objects}} & \multicolumn{1}{c||}{\textbf{Total}} & \multicolumn{1}{c|}{{\bf Violations}} \\ \hline
	\hline

\multirow{2}{*}{\shortstack[l]{\tool{}\\(this work)}} & Benign                     & \multirow{2}{*}{Jan 2014 - Dec 2016} & 116,993                                  & 116,993 & \multirow{2}{*}{-}                                                       \\
	\cline{2-2}\cline{4-5}
                       & Malicious &  & 12,735 & 12,735 & \\
                       \hline
                       \hline

\multirow{2}{*}{\cite{Arp:Drebin}, \cite{Papernot:ESORICS}} & Benign                     & \multirow{2}{*}{Aug 2010 - Oct 2012} & 123,453                                  & 123,453 & \multirow{2}{*}{C1}                                                       \\
	\cline{2-2}\cline{4-5}
                       & Malicious &  & 5,560 & 5,560 & \\
                       \hline
                       \hline
\multirow{7}{*}{\cite{Mariconti:MaMaDroid}} & \multirow{2}{*}{Benign}    & Apr 2013 - Nov 2013 & 5,879                                    & \multirow{2}{*}{8,447}             &  \multirow{7}{*}{\shortstack{(C1)\\C2\\C3}} \\
                       &                            & Mar 2016            & 2,568                                    &                    &                  \\ \cline{2-5}
                       & \multirow{5}{*}{Malicious} & Oct 2010 - Aug 2012 & 5,560                                    & \multirow{5}{*}{35,493}             & \\
                       &                            & Jan 2013 - Jun 2013 & 6,228                                    &                                     & \\
                       &                            & Jun 2013 - Mar 2014 & 15,417                                   &                                     & \\
                       &                            & Jan 2015 - Jun 2015 & 5,314                                    &                                     & \\
                       &                            & Jan 2016 - May 2016 & 2,974                                    &                                     & \\ \hline

\end{tabular}
}

	\caption{Summary of datasets composition used by our paper (1st row) and by prior work~\cite{Arp:Drebin,Mariconti:MaMaDroid,Papernot:ESORICS} (2nd and 3rd row).}

	\label{tab:datasets}
\end{table}

\end{appendix}

\end{document}